\documentclass[smus]{snow2e}
\usepackage{graphics,epsfig,rotating}
\def\nn{\noindent}
\def\Re{{\cal R \mskip-4mu \lower.1ex \hbox{\it e}\,}}
\def\Im{{\cal I \mskip-5mu \lower.1ex \hbox{\it m}\,}}
\def\ie{{\it i.e.}}
\def\eg{{\it e.g.}}

\def\etal{{\it et al.}}

\def\sub#1{_{\lower.25ex\hbox{$\scriptstyle#1$}}}
\def\tev{\,{\ifmmode\mathrm {TeV}\else TeV\fi}}
\def\gev{\,{\ifmmode\mathrm {GeV}\else GeV\fi}}
\def\mev{\,{\ifmmode\mathrm {MeV}\else MeV\fi}}
\def\to{\rightarrow}

\def\lum{\ifmmode {\cal L}\else ${\cal L}$\fi}
\def\inpb{\ifmmode {\rm pb}^{-1}\else ${\rm pb}^{-1}$\fi}
\def\infb{\ifmmode {\rm fb}^{-1}\else ${\rm fb}^{-1}$\fi}
\def\epem{\ifmmode e^+e^-\else $e^+e^-$\fi}
\def\ppb{\ifmmode \bar pp\else $\bar pp$\fi}
\newskip\zatskip \zatskip=0pt plus0pt minus0pt
\def\matth{\mathsurround=0pt}
\def\lsim{\mathrel{\mathpalette\atversim<}}
\def\gsim{\mathrel{\mathpalette\atversim>}}
\def\atversim#1#2{\lower0.7ex\vbox{\baselineskip\zatskip\lineskip\zatskip
  \lineskiplimit 0pt\ialign{$\matth#1\hfil##\hfil$\crcr#2\crcr\sim\crcr}}}

\begin{document}
\onecolumn

\vbox{ \large
\begin{flushright}
SLAC-PUB-7440\\
OCIP/C 97-03\\
\end{flushright}

\vspace{1in}

\begin{center}
{\Large\bf Discovery Potential for New Phenomena$^*$}
\medskip
\vskip .3cm

{\large Stephen Godfrey$^{1}$, JoAnne L. Hewett$^{2}$, and 
Lawrence E. Price$^{3}$} \\
$^1${\em Ottawa Carleton Institute for Physics, 
Carleton University, Ottawa, Canada}\\
$^2${\em Stanford Linear Accelerator Center, Stanford, CA 94309, USA}\\
$^3${\em High Energy Physics Division, Argonne National Laboratory, Argonne, IL
60439, USA}\\
\vskip 1.0cm

{\bf Working Group Members}:
J. Appel (FNAL),
P. de Barbaro (Rochester),
M. Berger$^\dagger$ (Indiana),
G. Burdman (Wisconsin),
K. Cheung$^\dagger$ (Texas),
F. Cuypers (PSI),
S. Davidson (Max Planck),
M. Doncheski (Penn State - Mont Alto),
E. Eichten (FNAL),
C. Greub (DESY),
R. Harris$^\dagger$ (FNAL),
X.-G. He (Melbourne),
C. Heusch (U.C. Santa Cruz),
H. Kagan$^\dagger$ (Ohio State),
P. Kalyniak (Carleton),
D. Krakauer$^\dagger$ (ANL),
K. Kumar (Princeton),
T. Lee (FNAL),
J. Lykken (FNAL),
K. Maeshima$^\dagger$ (FNAL),
I. Melo (Carleton),
W. Merritt$^\dagger$ (FNAL),
P. Minkowski (PSI),
R. Peccei (UCLA),
S. Riemann (Zeuthen),
T. Rizzo$^\dagger$ (SLAC),
J. Rowe (U.C. Davis),
D. Silverman (U.C. Irvine),
E. Simmons (Boston),
J. Slaughter (FNAL),
M. Swartz$^\dagger$ (SLAC),
D. Toback (Chicago),
R. Vidal (FNAL),
J. Womersley (FNAL),
G. Wrochna (CERN),
J. Wudka (U.C. Riverside),
C.-E. Wulz (Austria, OAW)

\vskip1cm

{\bf Abstract}\\
\end{center}

We examine the ability of future facilities to discover and interpret
non-supersymmetric new phenomena.  We first explore explicit manifestations
of new physics, including extended gauge sectors, leptoquarks, exotic fermions, 
and technicolor models.  We then take a more general approach where 
new physics only reveals itself through the existence of effective
interactions at lower energy scales.

\vspace*{0.5in}
\noindent Summary Report of the New Phenomena Working Group.
To appear in the {\it Proceedings of the 1996 DPF/DPB Summer
Study on New Directions for High Energy Physics - Snowmass96}, Snowmass, CO,
25 June - 12 July 1996.
\vskip 1.0in

\noindent $^*$ Work supported by the U.S. Department of Energy under contracts
DE-AC03-76SF00515 and W-31-109-ENG-38, and the Natural Sciences and 
Engineering Research Council of Canada\newline
\noindent $^\dagger$ Subgroup Convener

\thispagestyle{empty}
}
\newpage

\twocolumn

\title{Discovery Potential for New Phenomena}

\author{Stephen Godfrey$^{1}$, JoAnne L. Hewett$^{2}$, and 
Lawrence E. Price$^{3}$ \\
$^1${\em Ottawa Carleton Institute for Physics, 
Carleton University, Ottawa, Canada}\\
$^2${\em Stanford Linear Accelerator Center, Stanford, CA 94309, USA}\\
$^3${\em High Energy Physics Division, Argonne National Laboratory, Argonne, IL
60439, USA}\\
}

\maketitle

{\bf Working Group Members}\thanks{($^\dagger$ Subgroup Convener)}:
J. Appel (FNAL),
P. de Barbaro (Rochester),
M. Berger$^\dagger$ (Indiana),
G. Burdman (Wisconsin),
K. Cheung$^\dagger$ (Texas),
F. Cuypers (PSI),
S. Davidson (Max Planck),
M. Doncheski (Penn State - Mont Alto),
E. Eichten (FNAL),
C. Greub (DESY),
R. Harris$^\dagger$ (FNAL),
X.-G. He (Melbourne),
C. Heusch (U.C. Santa Cruz),
H. Kagan$^\dagger$ (Ohio State),
P. Kalyniak (Carleton),
D. Krakauer$^\dagger$ (ANL),
K. Kumar (Princeton),
T. Lee (FNAL),
J. Lykken (FNAL),
K. Maeshima$^\dagger$ (FNAL),
I. Melo (Carleton),
W. Merritt$^\dagger$ (FNAL),
P. Minkowski (PSI),
R. Peccei (UCLA),
S. Riemann (Zeuthen),
T. Rizzo$^\dagger$ (SLAC),
J. Rowe (U.C. Davis),
D. Silverman (U.C. Irvine),
E. Simmons (Boston),
J. Slaughter (FNAL),
M. Swartz$^\dagger$ (SLAC),
D. Toback (Chicago),
R. Vidal (FNAL),
J. Womersley (FNAL),
G. Wrochna (CERN),
J. Wudka (U.C. Riverside),
C.-E. Wulz (Austria, OAW)

\vspace*{0.4cm}

\thispagestyle{empty}\pagestyle{empty}

\begin{abstract}
We examine the ability of future facilities to discover and interpret
non-supersymmetric new phenomena.  We first explore explicit manifestations
of new physics, including extended gauge sectors, leptoquarks, exotic fermions, 
and technicolor models.  We then take a more general approach where 
new physics only reveals itself through the existence of effective
interactions at lower energy scales.
\end{abstract}

\section{Introduction}

Although the Standard Model (SM) of particle physics is in complete
agreement with present experimental data, it is believed to leave many
questions unanswered and this belief has resulted in numerous attempts to 
discover a more fundamental underlying theory.  In planning for the future, it 
is reasonable to consider what classes of new interactions might exist and 
what types of facilities would be best to first discover them and then to
elucidate their properties.  In fact, numerous types of experiments may
expose the existence of new physics; here we focus on the potential signatures 
at high energy colliders.  

History shows us that the most exciting discoveries
are those that are unexpected.  Unfortunately, it is difficult to concretely
plan for the unexpected.  The best we can do is to examine the discovery
capabilities of future facilities for a wide variety of anticipated particles 
under the hope that they will be sufficient in uncovering the truth in nature.
To accomplish this task, the 1996 Snowmass working group on new phenomena
decided to construct a physics matrix, where numerous new physics possibilities
were investigated at various collider options.  The accelerators used for our
physics studies were those defined by the Snowmass organizing committee.
The new phenomena scenarios were divided into three main categories: (i) New
Gauge Bosons, (ii) New Particles, and (iii) New Interactions.  The remainder 
of this report presents the conclusions from each category.  We note that our
physics matrix is strikingly similar to that presented in the Proceedings of
the 1982 Snowmass Summer Study\cite{snow}, both in physics topics and
colliders.  It is disappointing that so little progress has been made in 
our attempt to understand the fundamental theory of nature.

Before turning to our investigations of searches for new phenomena at high
energy colliders, we note that virtual effects of new physics also
provides an important opportunity to probe the presence of new 
interactions\cite{htt}.
This complementary approach examines the indirect effects of new physics in
higher order processes by testing for deviations from SM predictions.  In this
case, one probes (i) the radiative corrections to perturbatively calculable
processes, as well as (ii) transitions which are either suppressed or
forbidden in the SM.  Both of these scenarios carry the advantage of being able
to explore the presence of new physics at very high mass scales.  In some 
cases the constraints obtained in this manner surpass those from collider
searches, with a recent example being given by the strong bounds on the
mass of a charged Higgs boson from the decay $B\to X_s\gamma$\cite{bsg}.
In other cases, entire classes of models are found to be incompatible with
the data.  Given the large amount of high luminosity `low-energy' data which
is presently available and will continue to accumulate during the next decade,
the loop effects of new interactions in rare processes and precision
measurements will play a major role in the search for physics beyond the SM.

\begin{figure}[h]
\leavevmode
\centerline{
\begin{turn}{-90}
\epsfig{file=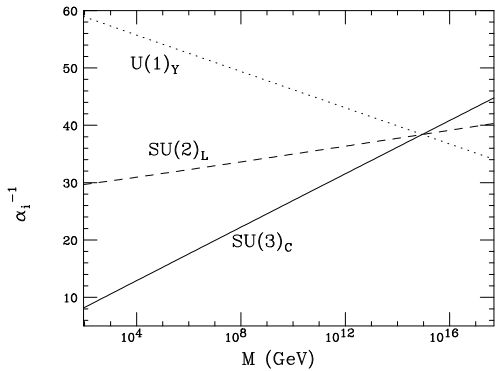,width=6.0cm,clip=}
\end{turn}
} 
\caption{Two-loop Renormalization Group Evolution of the coupling constants in
the scenario where the SM particle content is augmented by a pair of leptoquarks
and an additional Higgs doublet.  (From Ref. \cite{lqtj}.)}
\label{lqgut}
\end{figure}

It is well-known that physics outside of the SM is required in order to obtain
unification of the strong and electroweak forces.  Unification attempts
using only the SM particle content fail because they predict too small a
value of the unification scale, implying a rapidly decaying proton, as well as
leading to values of $\alpha_s(M_Z)$ which are significantly smaller than the
experimentally determined value by many standard deviations.  The oft-quoted
remedy to this situation is to introduce supersymmetry at the TeV 
scale\cite{amal,framp}.  
In fact, the introduction of the minimal supersymmetric 
particle content modifies the evolution of the coupling constants such that
unification is obtained at a higher scale and there is agreement with present
data.  The most frequently considered case is where supersymmetry (SUSY) is
embedded into a SUSY $SU(5)$ Grand Unified Theory (GUT).  However, satisfactory
unification is also achieved in larger SUSY GUTs, such as supersymmetric
$SO(10)$ and $E_6$.  In these cases both the gauge sector and particle content 
are enlarged, leading to the many possible types of new phenomena which are
discussed in the first two sections of this report.  In particular, it has been 
shown\cite{deshtgr} that successful unification is achievable in SUSY $SO(10)$
with a light right-handed mass scale, resulting (amongst other things) in a 
right-handed $W$-boson which would be accessible to experiment.  It
is also possible for non-supersymmetric models with additional particle
content to show similar unification properties\cite{framp,tgrguts}, although
such cases are difficult to arrange.  One such scenario is that where the
SM particle content is augmented by a weak iso-doublet of leptoquarks 
($\widetilde R_{2L}$, to be defined below) and an additional Higgs doublet.  
The two-loop renormalization group analysis of this case\cite{lqtj} is 
presented in Fig. \ref{lqgut}.  Here, one obtains the value
$\alpha_s(M_Z)=0.123$ and a proton lifetime of $10^{32\pm 1}$ years consistent
with experiment.

\section{New Gauge Bosons}

New gauge bosons (NGBs) are a feature of many extensions of the standard 
model such as grand unified theories, Left-Right symmetric models, 
supersymmetric models, and superstring theories.  If a $Z'$ or $W'$ 
were discovered it would have important implications for what lies 
beyond the standard model.  It is therefore important to study and 
compare what the next generation of colliders can tell us about 
NGBs.   There is a vast literature on the subject of discovery and 
identification of NGBs \cite{c-v}.  The NGB subgroup goals were to 
extend previous studies in several directions:
\begin{enumerate}
\item To extend the existing analysis to the colliders included 
as part of the Snowmass study that have not been previously studied.  
\item To extend studies to include gauge bosons that have received 
incomplete attention in the past,  in particular discovery reaches for $W'$ 
bosons at $e^+e^-$ colliders.  
\item To redo earlier studies including important 
considerations so far neglected.  For example, the cross sections for 
$Z'$ and $W'$'s at hadron colliders have almost always been calculated 
in the narrow width approximation, generally decaying only to standard 
model fermions, and not including backgrounds.  However, once the 
widths are included the signal may broaden sufficiently that it is 
overwhelmed by background.  At the minimum, it is important to know 
how these contributions  will effect the discovery limits.
\end{enumerate}
In this report we summarize the results of these studies and attempt 
to integrate them with previous results to present a complete overview 
of the subject of NGBs.  By necessity this summary will omit important
details of the various calculations.  We therefore direct the interested 
reader to the more complete and 
detailed subgroup summary by Rizzo \cite{ngb-sum} 
and the individual contributed reports to the proceedings.

\subsection{Introduction to Models}

Quite a few models predicting NGBs exist in the literature.  These 
can be divided into two broad classes depending 
on whether or not they originate from a GUT group such as $SO(10)$ or 
$E_6$.  We focus our studies on a few representative models, which 
although far from exhaustive, form a representative set for the 
purposes of this study.  To be specific the models we consider are:
\begin{enumerate}
\item The $E_6$ effective rank-5 model (ER5M) which predicts a $Z'$ 
whose couplings depend on a parameter $-\pi/2 \leq \theta_{E_6} \leq 
\pi/2$. Models $\psi$ $(\theta_{E_6} =0)$, $\chi$ $(\theta_{E_6} 
=-\pi/2)$, I $(\theta_{E_6} =-\cos^{-1} \sqrt{3/8})$,
and $\eta$ $(\theta_{E_6} =\cos^{-1} \sqrt{5/8})$ denote common cases 
discussed in the literature.
\item The Sequential Standard Model (SSM) where the new $W'$ and $Z'$ 
are just heavy versions of the SM.  This is not a true model but is 
often used as benchmark by experimenters.
\item The Un-Unified Model (UUM) based on the group $SU(2)_\ell \times 
SU(2)_q \times U(1)_Y$, which has a single free parameter $0.24 \leq 
s_\phi \leq 0.99$
\item The Left-Right Symmetric Model (LRM) based on the group 
$SU(2)_L \times SU(2)_R \times U(1)_{B-L}$ which has a free parameter 
$\kappa =g_R/g_L \geq 0.55$ which is just the ratio of the gauge 
couplings.
\item The Alternative Left-Right Model (ALRM) based on the same 
extended group as the LRM but now arising from $E_6$ where the fermion 
assignments are different from those in the LRM due to an ambiguity in 
how they are embedded in the {\bf 27} representation.
\end{enumerate}
Details of these models and complete references are given in Ref. \cite{c-v}. 

Although searches for NGBs, and indeed any new particles, are of 
interest on general grounds,  if there are theoretical motivations for 
them to be accessible at existing or future colliders their 
phenomenological interest is enhanced considerably.  In a contribution 
to these proceedings, Lykken \cite{lykken} examined this issue for the 
case of a new $U(1)'$ gauge group in the general context of SUSY-GUTS 
and String Theory with weak-scale supersymmetry.  He found that a broad
class of models predict a $Z'$ boson whose mass is in the range
$250\gev-2\tev$.  However, these models require either discrete tuning
of the $U(1)'$ charges or a leptophobic $Z'$.

\subsection{Discovery Limits}

Two distinct search strategies exist for extra gauge bosons.  Indirect 
evidence  for gauge bosons, where deviations from the standard model would 
signal new physics, are the primary approach at $e^+e^-$, $e^-e^-$, 
$\mu^+ \mu^-$, and $ep$ 
colliders while direct evidence signalled but clusters of high 
invariant mass lepton pairs is the primary strategy employed at hadron 
colliders.  A large literature exists on search strategies for extra 
gauge bosons and their discovery limits, for existing and 
proposed high energy colliders.  

\subsubsection{Hadron Colliders}

In hadron colliders NGBs will generally reveal themselves through 
decays to charged lepton pairs for $Z'$ bosons and to charged leptons plus 
missing $E_T$ for $W'$ bosons.  There are exceptions such as leptophobic 
$Z'$ bosons decay to quark pairs which would be observed as bumps 
in dijet invariant mass distributions \cite{toback}.  

Search limits have been obtained previously for all the hadron 
colliders  \cite{godfrey} 
considered at Snowmass with the exception of the 200~TeV 
(PIPETRON) collider.  
However, these results were generally obtained 
using the narrow width approximation with the NGB decaying only to 
conventional fermions and with possible corrections to 
account for detector acceptances and efficiencies.  Discovery was 
defined to be 10 dilepton signal events.  Detailed detector 
simulations for the the Tevatron and LHC validated this approximation 
as a good estimator of the true search reach.  The discovery reaches 
for hadron colliders are summarized in Table I \cite{rizzo-zp}.
TeV33 will, for the first time allow us to approach the 1 TeV mass 
scale for $Z'$ bosons.  For the 60 and 200 TeV machines the higher 
$q\bar{q}$ luminosities in the $p\bar{p}$ mode leads to significantly 
greater $(\simeq 30 - 50\%)$ search reach. 
It is important to note 
that in many models the $Z'$ can also decay to exotic fermions and/or 
SUSY particles which will reduce $B_\ell$ and thus the search reach 
(about 10\% reduction in search reach
for a factor of 2 decrease in $B_\ell$) \cite{rizzo-zp,capstick}.  

\begin{figure}[h]
\leavevmode
\centerline{\epsfig{file=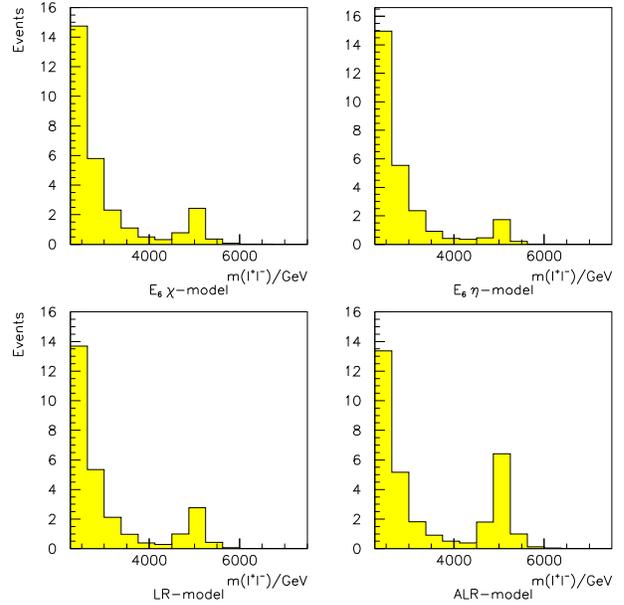,width=9.0cm,clip=}}
\caption{Invariant mass spectra for four $Z'$ models with 
$M_{Z'}=5$~TeV. (From Ref. \cite{wulz}.)}
\label{zppythia}
\end{figure}

Wulz performed detailed Monte Carlo studies of $Z'$ discovery limits 
for the LHC at $\sqrt{s}=14$~TeV 
using the CMS detector simulation  and PYTHIA to generate the $Z'$ 
events \cite{wulz}.  Exotic fermions were not assumed.  Drell-Yan and 
$Z$ backgrounds were taken into account and were approximately two 
orders of magnitude below the signal.  Heavy flavor backgrounds from 
$t\bar{t}$ and $b\bar{b}$ are completely negligible.  Figure \ref{zppythia} 
shows reconstructed invariant mass spectra for $M_{Z'}=5$~TeV and an 
integrated luminosity of 100~fb$^{-1}$.  The discovery limits obtained 
by Wulz are consistent with the numbers given in Table I.

Unlike the $Z'$ case, $W_R$ searches have many subtleties.  Typically, 
search limits are obtained by assuming (i) the $q'\bar{q} W_R$ 
production vertex has SM strength, (ii) $\kappa=1$, 
(iii) $|V_{L_{ij}}|=|V_{R_{ij}}|$ (the CKM mixing matrix $V_R\equiv V_L$), 
and (iv) $B(W_R \to \ell \nu)$ is given by the decay to SM fermions. If 
assumption (ii) is invalid large search reach degradations are 
possible, especially at $p\bar{p}$ colliders, due to modification of 
the parton luminosities \cite{rizzo-zp}.  Again, the search reaches 
are higher $(\sim 25\%)$ in the case of $p\bar{p}$.

\subsubsection{Lepton Colliders}

If $Z'$'s and $W'$'s are to be found at lepton colliders their 
existence is most likely to be revealed
through deviations from SM predictions.  To 
represent a meaningful signal of new physics deviations should be 
observed in as many observables as possible.  Typically observables are 
constructed from cross sections to specific fermions in the final 
state; cross sections, $\sigma^f$, 
forward-backward asymmetries, $A_{FB}^f$, 
and left-right polarization asymmetries, $A_{LR}^f$,
where $f=\mu$, $\tau$, $c$, $b$, and $had=$sum over hadrons.  
Expressions for these observables are included in the contribution of 
Godfrey \cite{godfrey-sm}.  
The report by Godfrey gives discovery limits for high energy $e^+e^-$ and 
$\mu^+\mu^-$ colliders.  The main distinction between the two types of 
colliders is that $e^+e^-$ colliders should have high polarizations 
while $\mu^+\mu^-$ colliders are not expected to.  
That analysis assumed 90\% electron polarization (for the $e^+e^-$ 
case), 35\% $c$-tagging efficiency and 60\% $b$-tagging efficiency.  
In retrospect these efficiencies are likely to be overly optimistic 
for the $\mu^+\mu^-$ collider.  Rizzo performed a similar analysis 
except for the $e^+e^-$ colliders he included $t$-quark final states 
and the additional complications 
of angular cuts and initial state radiation(ISR) \cite{rizzo-zp}.  
He found that ISR reduces the search reach by 15-20\% while beam 
polarization increases the reach by 15-80\%, depending on the specific 
model and the machine energy.  

In principle the NLC can be run in a polarized $e^-e^-$ mode with
luminosity and polarization 
comparable to the $e^+e^-$ mode.  Since both $e^-$ beams are 
polarized the effective polarization is larger and, due to the large 
Moller cross section there is significant sensitivity to the existence 
of a $Z'$.  Cuypers studied the sensitivity of  $e^+e^-\to 
\mu^+\mu^-$,  $e^+e^-\to e^+e^-$ and $e^-e^-\to e^-e^-$ to $c_v'$ and 
$c_a'$ for fixed $M_{Z'}$ including ISR
and systematic errors due to imperfect polarization measurement, 
finite detector acceptance, and luminosity uncertainties  
\cite{cuypers-zp}. In general the $e^-e^-$ reach is slightly superior to that 
obtained in the $e^+e^-$ mode when only the leptonic final states are 
used.  However, Rizzo found that once 
one includes the additional information from the 
quark sector the $e^+e^-$ mode offers a higher reach \cite{rizzo-zp}. 

There has been very little work done on searches for $W'$ at $e^+e^-$ 
colliders.  In a contribution to the proceedings Hewett \cite{hewett} 
studied the sensitivity of the reaction $e^+e^- \to \nu\bar{\nu} 
\gamma$ to $W'$ bosons which would contribute via t-channel exchange.  She 
found that the resulting photon energy spectrum would be sensitive to 
a $W_R$ mass of at most $2\times \sqrt{s}$ in the LRM  and in the UUM 
the $W_H$ discovery reach barely extends above $\sqrt{s}$ for small 
values of $\sin \phi$.  However for larger values of $\sin \phi$ the 
reach grows to several times $\sqrt{s}$ due to the increase in 
leptonic couplings for $\sin \phi >0.5$.  Although these preliminary 
results do not directly compete with the discovery reach at the LHC 
they do demonstrate that it is possible to observe the effects of 
$W'$ bosons with masses greater than $\sqrt{s}$ at $e^+ e^-$ colliders.

\begin{table*}
\caption{New gauge boson search reaches in TeV.  
For the LRM $\kappa=1$ is assumed,
while for the UUM we take $s_\phi=0.5$.  Decays to SM fermions only
are taken into account.  The luminosities for the Tevatron, LHC, 60~TeV, and 
200~TeV colliders are taken to be 10, 100, 100, 1000~$fb^{-1}$, respectively.}
\begin{center}
\begin{tabular}{lcccccccc}
\hline
\hline
Machine & $\chi$ & $\psi$ & $\eta$ & SSM & LRM & ALRM & UUM & $W'$ \\
\hline
\hline
\multicolumn{9}{c}{Hadron Colliders} \\
\hline
CDF/D0 & 0.585 & 0.580 & 0.610 &  0.620 & 0.690 & --- & --- & 0.720 \\
Tev33 & 1.0 & 1.0 & 1.0 & 1.1 & 1.1 & 1.2 & 1.1 & 1.2 \\
LHC   & 4.6 & 4.1 & 4.2 & 4.9 & 4.5 & 5.2 & 4.6 & 5.9 \\
60 TeV $(pp)$ & 13 & 12 & 12 & 14 & 14 & 15 & 14 & 20 \\
60 TeV $(p\bar{p})$ & 18 & 17 & 18 & 21 & 19 & 22 & 20 & 25 \\
200 TeV $(pp)$	& 44 & 39 & 40 & 46 & 43 & 50 & 44 & 65 \\
200 TeV $(p\bar{p})$ & 64 & 62 & 65 & 69 & 65 & 75 & 65	& 83 \\
\hline
\multicolumn{9}{c}{Lepton Colliders} \\
\hline
NLC500 & 3.2 & 1.9 & 2.3 & 4.0 & 3.7 & 3.8 & 4.8 & \\
NLC1000 & 5.5 & 3.2 & 4.0 & 6.8 & 6.3 & 6.7 & 8.2 & \\
NLC1500 & 8.0 & 4.8 & 5.8 & 10 & 9.2 & 9.8 & 12 & \\
NLC 5 TeV & 23 & 14 & 17 & 30 & 26 & 28 & 35 & \\
$\mu^+\mu^-$ 4 TeV & 18 & 11 & 13 & 23 & 20 & 22 & 27 & \\
\hline
\hline
\end{tabular}
\end{center}
\end{table*}

\subsection{Coupling Determination}

If evidence for NGBs were found the next task would be to obtain 
information that would verify the discovery and 
could determine the nature of the NGB.  Both hadron and lepton 
colliders can play complementary roles in this task, each having 
strengths and weaknesses.
If a strong signal for a NGB
were obtained at hadron colliders one could proceed directly to 
measuring its couplings.  However, if the only evidence for NGBs came 
from lepton colliders, where the evidence is indirect, determining the 
nature of the new physics is more complicated.  To measure the 
couplings one would have to independently determine the NGB mass 
since coupling's values scale as $M_{Z'}^{-4}$.  
The recent review by Cvetic and Godfrey 
\cite{c-v} summarizes the current status of NGB identification.
For the most part existing studies of NGB identification have not 
included the limitations of using realistic detectors.  This is 
especially important for hadron colliders.  For lepton colliders 
virtually all existing studies examine how well couplings can be 
determined if the NGB mass is known.  A main effort of the
NGB subgroup was to extend existing
studies to realistic detectors and to determining how well NGB 
properties could be determined in a model {\em blind} approach.

\subsubsection{Hadron Colliders}

\begin{figure}[h]
\leavevmode
\centerline{\epsfig{file=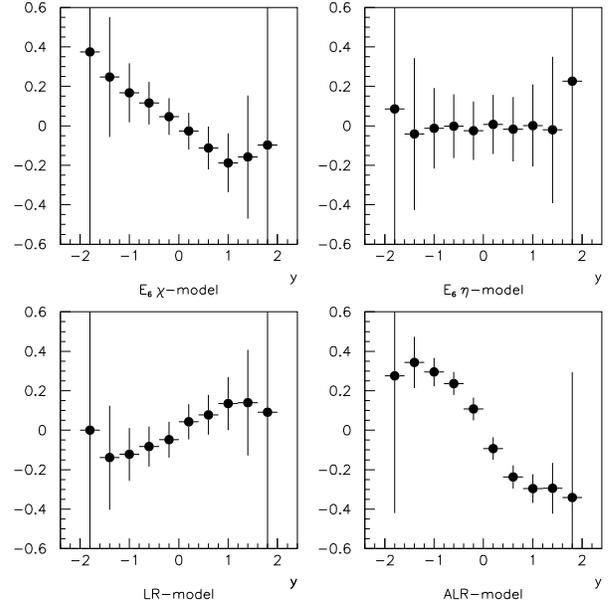,width=9.0cm,clip=}}
\caption{$Z'$ asymmetries in dilepton channels at the LHC
for $M_{Z'}=2$~TeV.  (From Ref. \cite{wulz}.)}
\label{zpasym}
\end{figure}

Although the total $Z'$ production cross section at a hadron collider
is a function of the $Z'$ 
couplings, the leptonic cross section depends on unknown contributions 
from supersymmetric particles or exotic fermions to the total width 
which makes its use as a tool to distinguish models questionable at best.  
The determination of $Z'$ couplings is a daunting task due to large 
backgrounds and limited statistics. 
The most widely used observable for model identification at 
hadron colliders is the forward backward asymmetry, $A_{FB}$.  Wulz 
examined this observable using rapidity bins.  The results for the LHC
are plotted for 
several models in Fig. \ref{zpasym}. It is clear that these models would be 
distinguishable for $M_{Z'}$ up to about 3~TeV and depending on the model,
information could be obtained up to about 3~TeV.  However, it is not 
clear as to what level one could extract precise coupling information.

\subsubsection{Lepton Colliders}

\begin{figure}[h]
\leavevmode
\begin{minipage}[t]{4.0cm}
\centerline{\epsfig{file=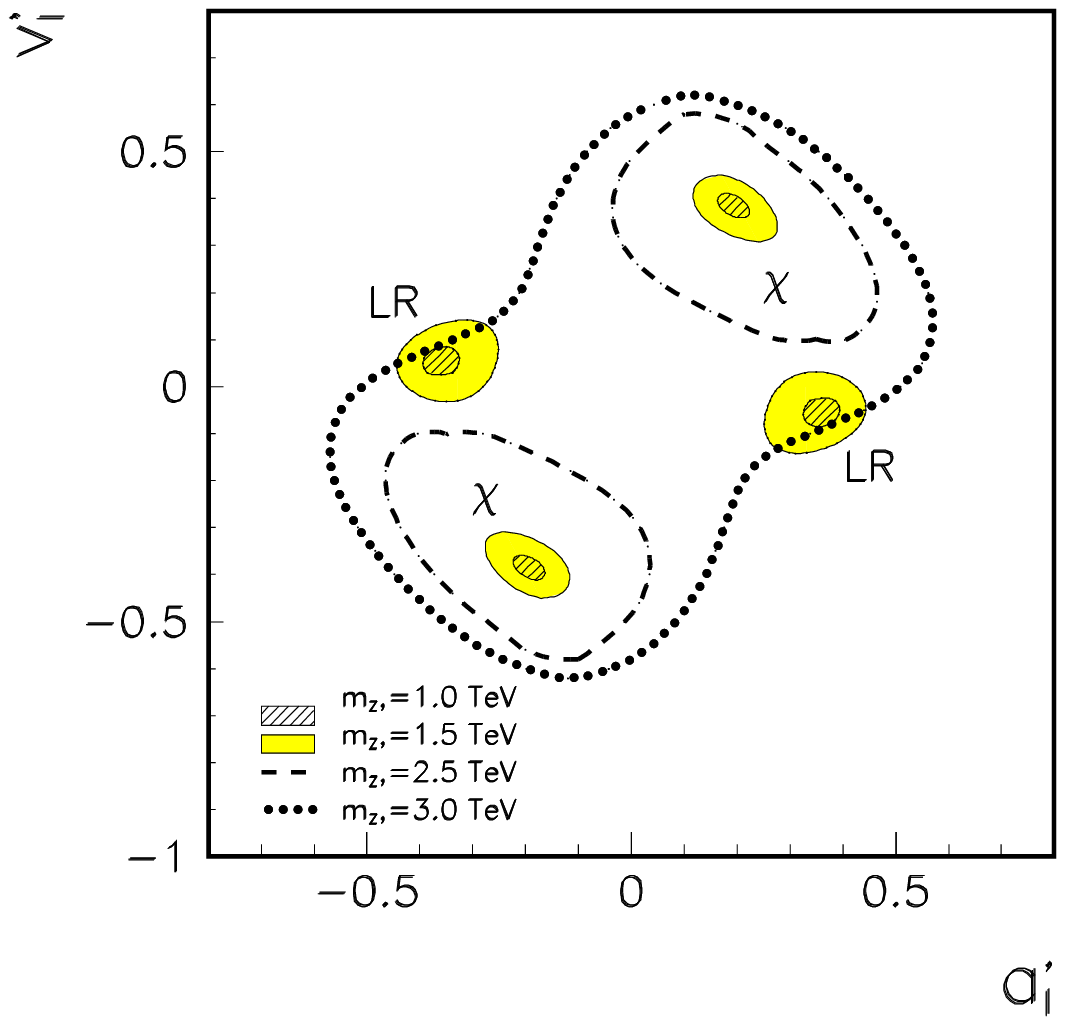,width=4.0cm,clip=}}
\end{minipage} \
\begin{minipage}[t]{4.0cm}
\centerline{\epsfig{file=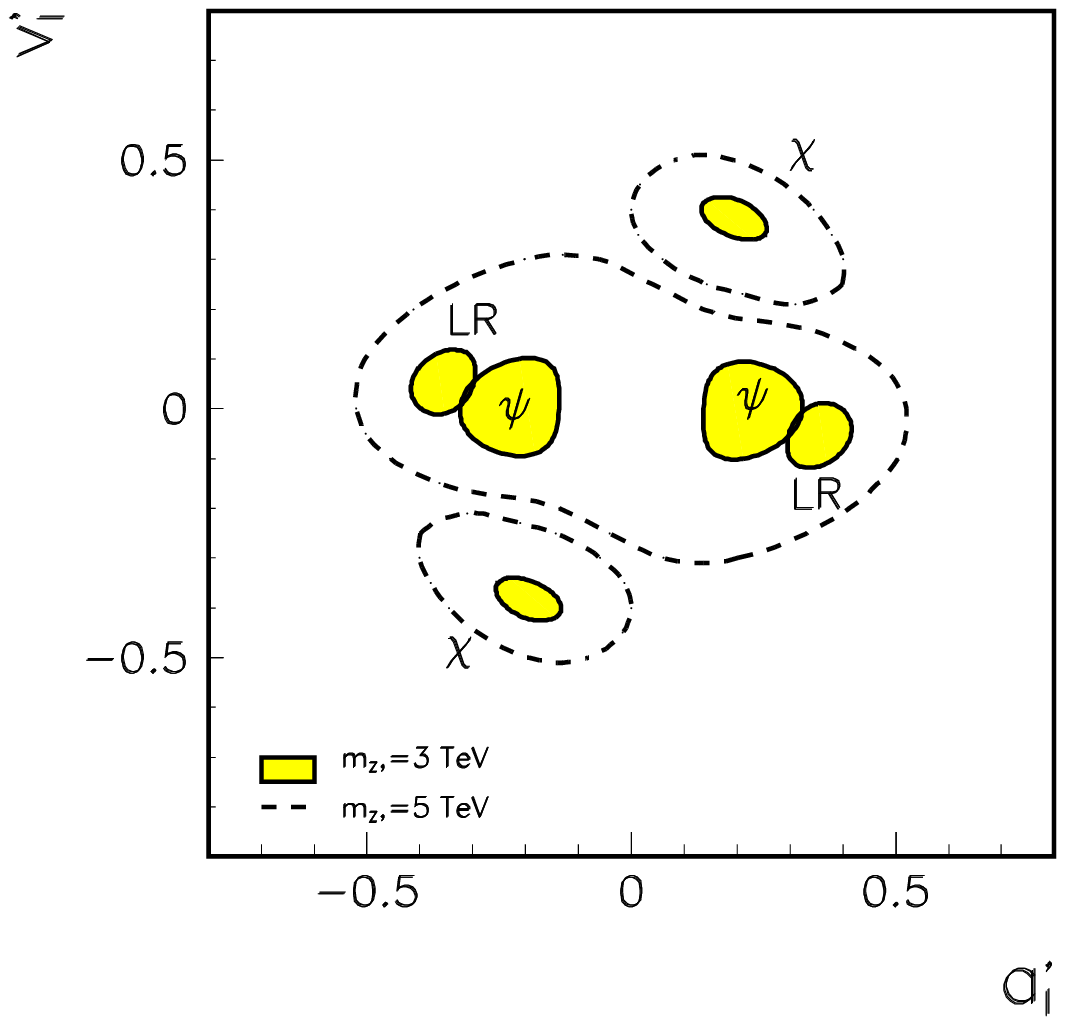,width=4.0cm,clip=}}
\end{minipage}
\caption{95\% C.L. contours for $a_\ell'$ and $v_\ell'$.  A $Z'$  is
assumed in the $\chi$, $\psi$, or LR model for different $Z'$ masses.
The left and right figures are
for $\sqrt{s}=500$~GeV, $L=50$~fb$^{-1}$
and $\sqrt{s}=1.5$~TeV, $L=100$~fb$^{-1}$. (From Ref. 
\cite{riemann-zp}.)}
\label{sabineone}
\end{figure}

If either evidence for NGBs were observed at a hadron collider or 
deviations from the SM that could be interpreted as a 
NGB were observed at a Lepton collider, the measurement of the NGB 
couplings would be of primary importance.  A number of contributions 
examined this problem for the NLC.  
Both Cuypers \cite{cuypers-zp} and 
Riemann \cite{riemann-zp} assumed a specific $Z'$ mass with the 
collider operating below this energy.  In his analysis Cuypers 
\cite{cuypers-zp}
included polarization error, detector angular resolution, initial 
state radiation, and luminosity measurement errors.  He assumed 
generic $v_{Z'}$ and $a_{Z'}$ couplings normalized to the charge of 
the electron $e$.  For a 
$\sqrt{s}=500$~GeV $e^+e^-$ collider with $L=50$~fb$^{-1}$ operating in 
either $e^-e^+$ or $e^-e^-$ mode, and assuming
$M_{Z'}=2$~TeV he found that $v_{Z'}$ and $a_{Z'}$ could be measured 
to about $\pm 0.3$.  
Riemann \cite{riemann-zp} followed a similar approach but presented 
her results in terms of the couplings for specific NGB models and how 
well they could be discriminated.  Riemann considered the NLC options;
$\sqrt{s}=500$~GeV $L=50$~fb$^{-1}$, 
$\sqrt{s}=1$~TeV $L=100$~fb$^{-1}$, and
$\sqrt{s}=1.5$~TeV $L=100$~fb$^{-1}$, with 80\% electron polarization, 
detector angular acceptances, quark flavor tagging efficiencies,  and 
luminosity measurement uncertainty of 0.05\%.  Riemann's results are 
summarized in Figure \ref{sabineone}. 
It is clear from these results that the NLC will be able to extract 
leptonic coupling information for $Z'$ masses up to $2-3\sqrt{s}$.
It should be noted that the lepton 
observables only depend on products or squares of $a_f'$ and $v_f'$ 
which results in a two-fold ambiguity in the signs of the couplings.  

\begin{figure}[b]
\leavevmode
\begin{minipage}[t]{4.0cm}
\centerline{\epsfig{file=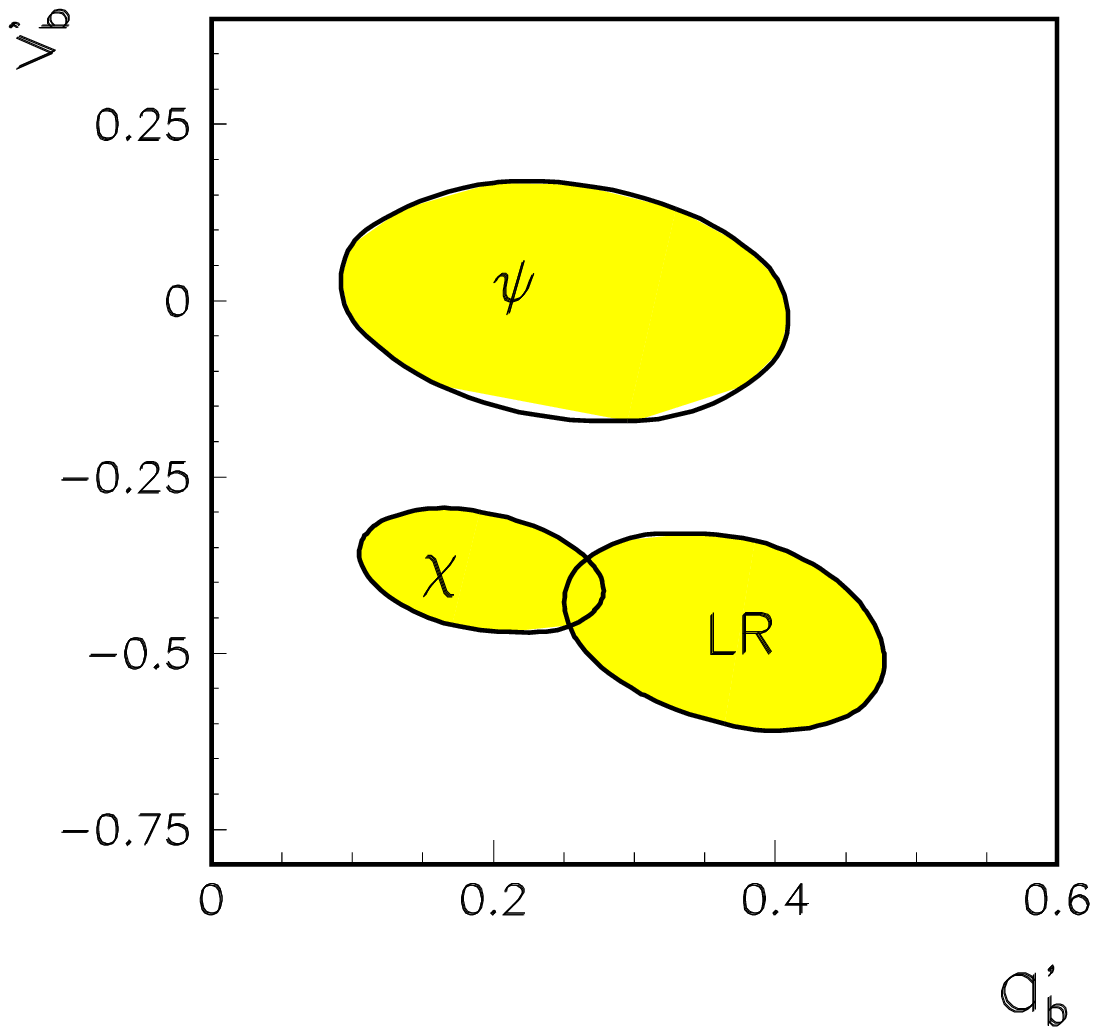,width=4.0cm,clip=}} 
\end{minipage} \
\begin{minipage}[t]{4.0cm}
\centerline{\epsfig{file=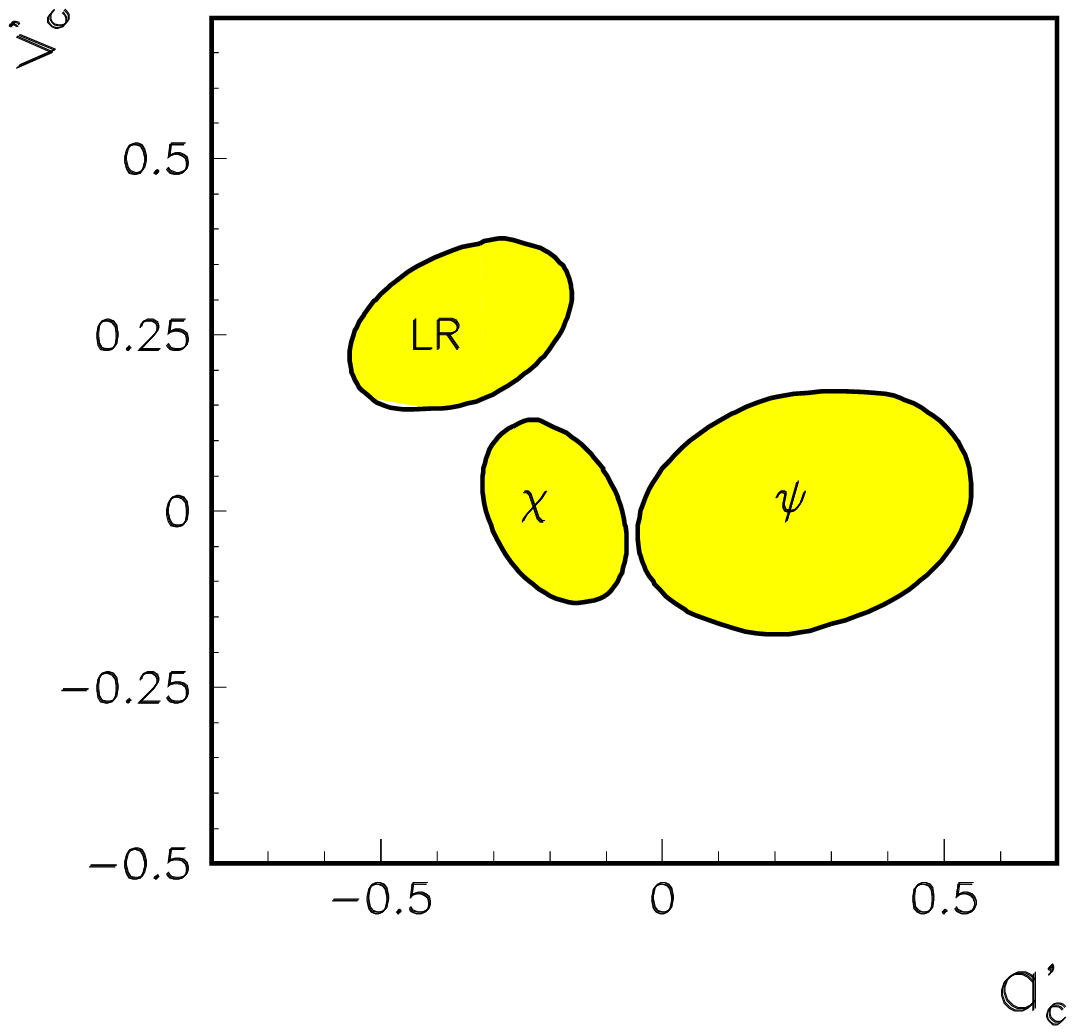,width=4.0cm,clip=}}
\end{minipage}
\caption{Model discrimination for $M_{Z'}=1$~TeV at $\sqrt{s}=0.5$~TeV 
with L=50~fb$^{-1}$
for $e^+e^-\to b\bar{b}$ (left) and $e^+e^-\to c\bar{c}$ (right). 
60\% (40\%) tagging efficiencies and 1\% (1.5\%) systematic errors 
were used for $b$ ($c$).  (From Ref. \cite{riemann-zp}.)}
\label{sabinetwo}
\end{figure}

Riemann also studied model discrimination using heavy flavor 
tagging.  The expected results for the $\sqrt{s}=500$~GeV collider with 
$M_{Z'}=1$~TeV are shown in Fig. \ref{sabinetwo}.  
Riemann stresses that these results are sensitive to the systematic 
errors for the measurements on these final states.

\begin{figure}[b]
\leavevmode
\begin{minipage}[b]{4.0cm}
\centerline{
\begin{turn}{-90}
\epsfig{file=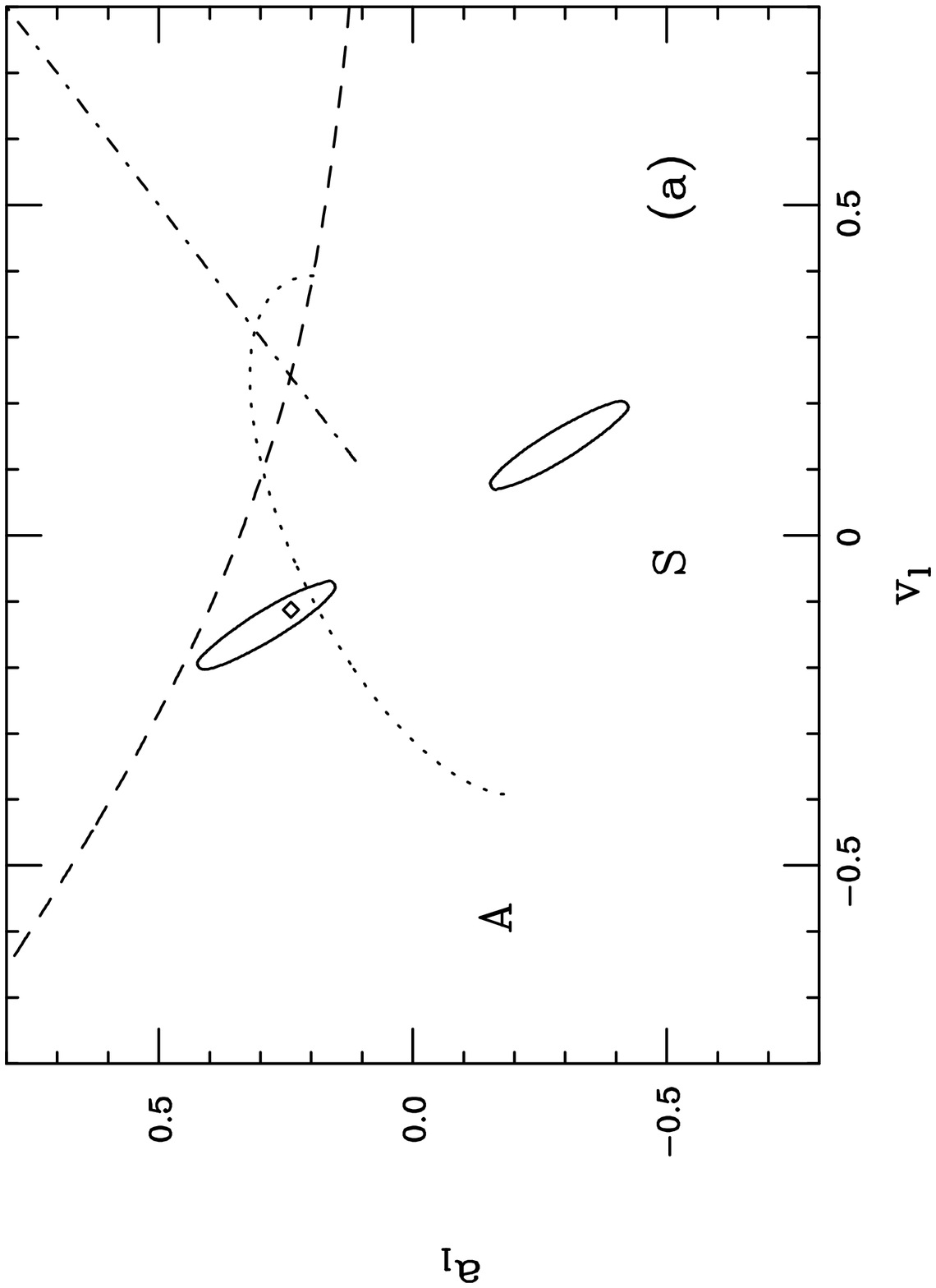,width=4.0cm,clip=}
\end{turn}
} 
\end{minipage} \
\begin{minipage}[b]{4.0cm}
\centerline{
\begin{turn}{-90}
\epsfig{file=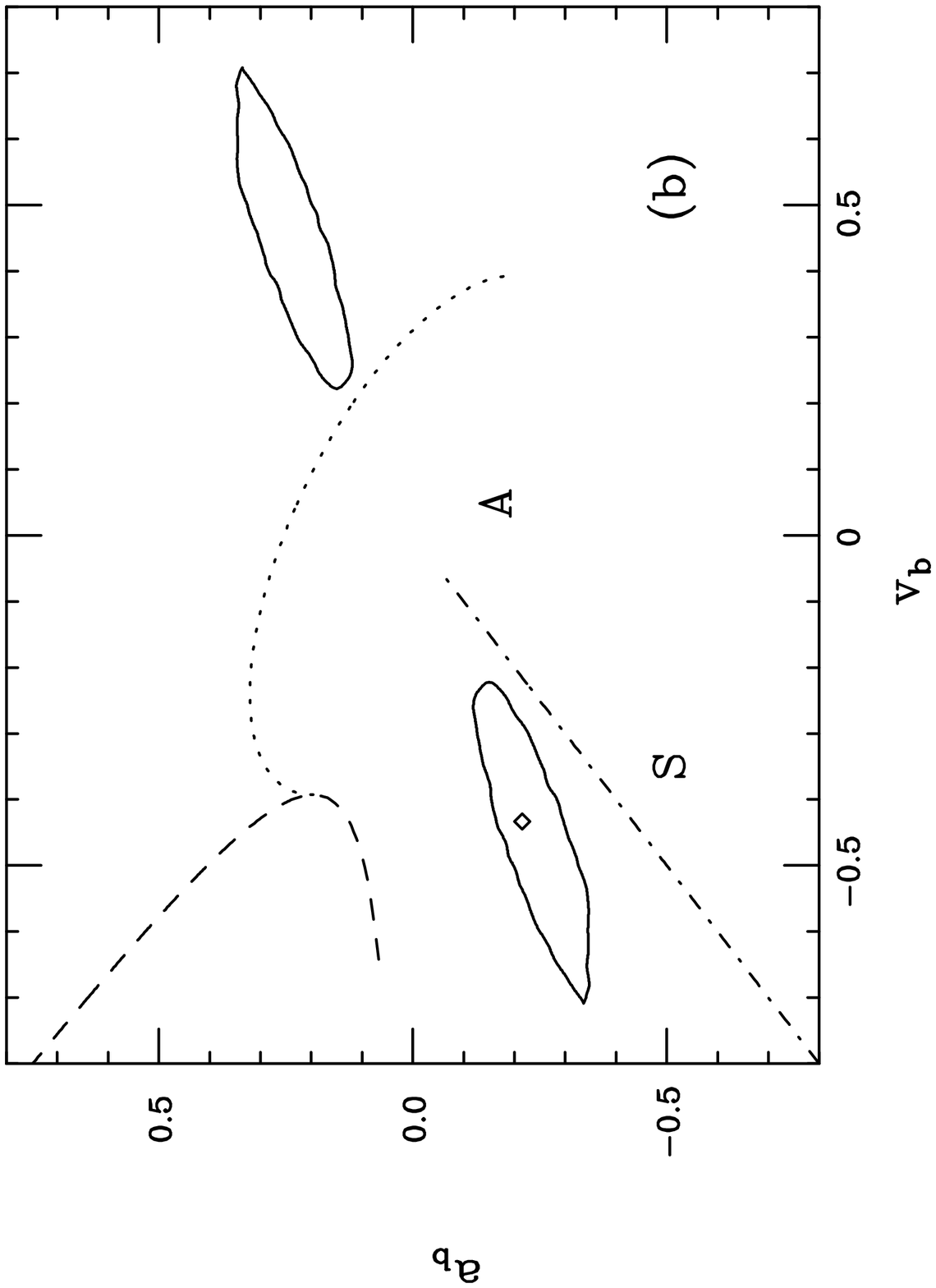,width=4.0cm,clip=}
\end{turn}
}
\end{minipage}
\centerline{
\begin{turn}{-90}
\epsfig{file=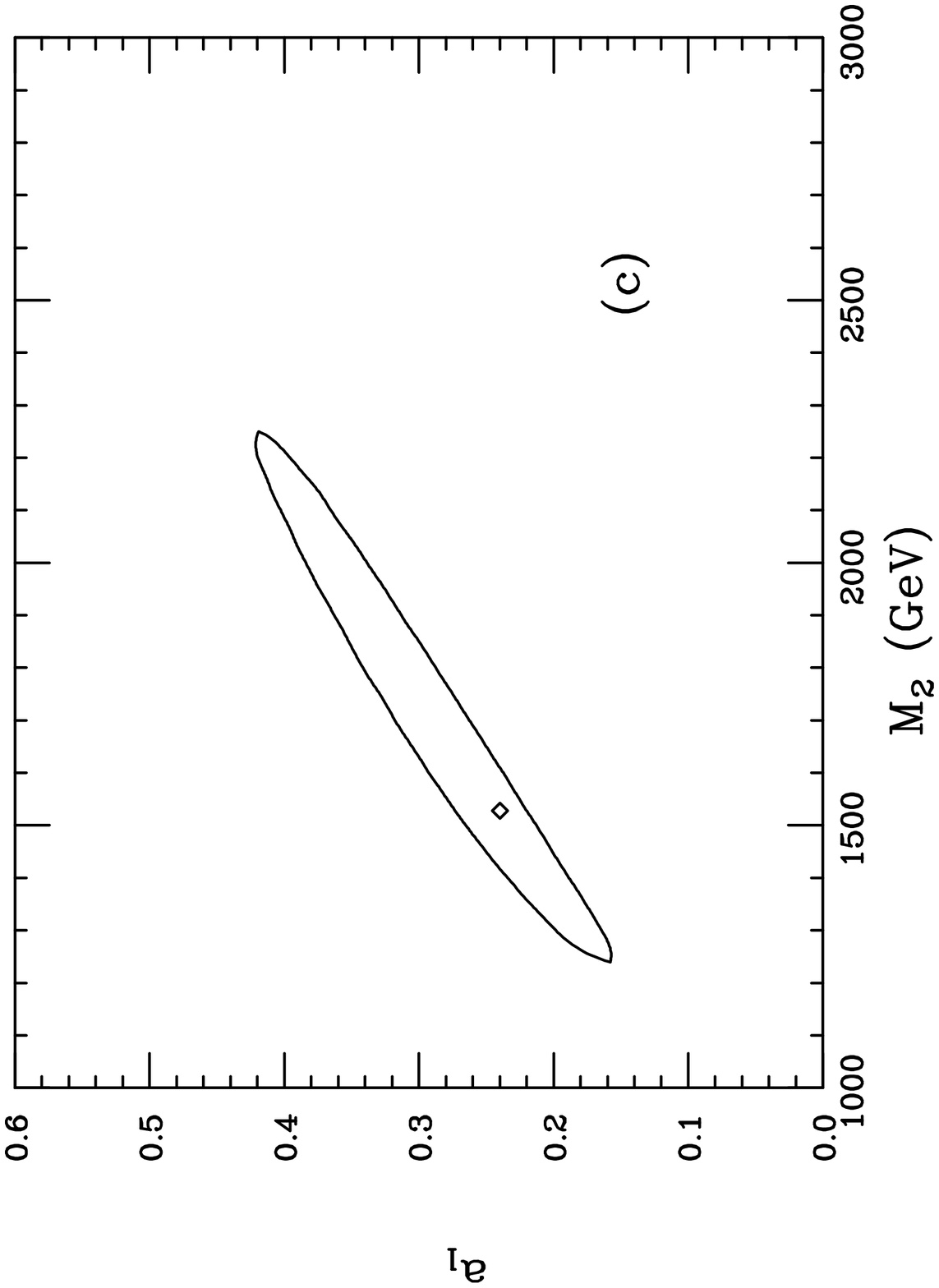,width=4.0cm,clip=}
\end{turn}
} 
\caption{95\% CL allowed regions for the extracted values of the (a) 
lepton (b) $b$-quark couplings, and (c) $M_{Z'}$
for randomly selected  $Z'$ parameters 
compared to the predictions of the $E_6$ model (dotted), LR model 
(dashed), UUM (dash-dot), SSM (S), and ALR (A).  For (c) only the 
$a_\ell >0$ branch is shown.  In all cases the diamond represents the 
corresponding input values.}
\label{tgrzp}
\end{figure}

Rizzo examined the capabilities of the NLC to determine both the mass 
as well as the couplings to leptons and $b$-quarks of $Z'$'s below 
production threshold.  This can be done by collecting data at several 
different values of $\sqrt{s}$.  In his analysis he assumed 
$e,\mu,\tau$ universality, 90\% $e^-$ polarization, 50\% $b$-tagging 
efficiency, 0.25\% luminosity measurement error, angular detector 
acceptance cut of $|\theta|>10^o$, final state QED and QCD corrections 
are included,  and neglecting t-channel exchange in $e^+e^-\to e^+e^-$.  
To insure model-independence the values of the $Z'$ couplings 
($v_{\ell,b},a_{\ell,b}$) and $M_{Z'}$ were chosen randomly and 
anonymously. Performing the analysis for a wide range of possible mass 
and couplings then shows the power as well as the limitations of the 
technique.  The results of one such analysis are shown if Fig. \ref{tgrzp}
where data was generated for $\sqrt{s}=0.5$ 0.75, and 1~TeV with 
associated integrated luminosities of 70, 100, and 150~fb$^{-1}$. 
A 5-dimensional 95\% C.L. allowed region for the mass and couplings is 
then found from a simultaneous fit of the various observables for the 
given energies.  Figure \ref{tgrzp} shows projections of the 5-dimensional 
region onto a 2-dimensional plane.  To give these results a context, the 
expectations of several well-known $Z'$ models are also shown.  
Rizzo's results again show the 2-fold ambiguity pointed out above.  
These results show that obtaining coupling information from different 
fermion species is important for discriminating between models. Rizzo 
also found that one needs at least 3 values of $\sqrt{s}$ to find 
$M_{Z'}$ and that spreading the integrated luminosities over too many 
Center of mass energies is also a failed strategy.  A final note is that 
previous knowledge of the value of
$M_{Z'}$ results in a much better measurement of the couplings.

\section{New Particles}

Many theories beyond the SM of the electroweak and strong interactions 
predict the existence of new particles. For the purposes of this report, 
these new states can be organized 
into two major categories: exotic fermions and difermions. Other new particle 
possibilities consist of new gauge bosons and excited fermions; these are 
discussed elsewhere. For a broad overview and introduction 
to new particles, as well as original references, see {\cite {dpf}} 
and the subgroup report by Berger and Merritt \cite{b-m}.

\subsection{Overview}

a) {\it Exotic Fermions.}  New fermions are predicted by many gauge 
extensions of the SM and they often have the usual lepton and baryon 
number assignments while possessing
non--canonical $SU(2)_L \times U(1)_Y$ quantum 
numbers, \eg, the left--handed components are in weak isosinglets and/or 
the right--handed components in weak isodoublets. Fourth generation 
fermions are sometimes considered in this class although their quantum numbers 
are canonical. Some examples of these exotic fermions are as follows: 

$i)$ Vector fermions: These are present, for instance, in $E_6$ grand unified
theories{\cite {pr}}.
In this example, each fermion generation lies in the representation of 
dimension {\bf 27}, and 
in addition to the fifteen SM chiral fields, twelve new fields are needed to 
complete this representation. Among these, there will be two weak isodoublets
of heavy leptons, one being right-handed and the other left-handed.  Vector 
fermions can have SM invariant masses and hence
contribute very little to the oblique parameters which describe
electroweak precision measurements\cite{htt}.

$ii)$ Mirror fermions: These have chiral properties which are opposite to 
those of ordinary fermions, i.e., the right-handed components are weak 
isodoublets and the left-handed ones are weak isosinglets.  There is also a 
left-handed heavy neutrino. These fermions appear in many extensions 
of the SM and provide a possible way to restore left--right symmetry at the 
scale of electroweak symmetry breaking. They have many of the 
phenomenological difficulties associated with fourth generation fermions,
such as the strict doublet mass splitting restrictions from contributions to
the $\rho$ parameter.

$iii$) Singlet fermions: These are the most discussed fermions in the 
literature, a prominent example being the right-handed neutrino in $SO(10)$.
Indeed, in this
unifying group, which is one of the simplest and most economic extensions of the
SM, the smallest anomaly free fermion representation has dimension {\bf 16}. It
contains the right-handed 
neutrino in addition to the 15 Weyl fermions in one fermion
generation; with this neutrino being 
of the Majorana type. Singlet neutrinos, which can
be either of Majorana or Dirac type, and new singlet quarks also occur in 
$E_6$ theories.

b) {\it Difermions.} These are scalar or vector particles which have unusual 
baryon and/or lepton number assignments. Examples of these particles are as 
follows:

$i)$ Leptoquarks: These particles are color triplets with B$=\pm 1/3$ and 
L$=\pm 1$. They naturally appear in models which place quarks and leptons on
an equal footing, such as
Technicolor, composite models (where quarks
and leptons are made of the same  subconstituents) as bound states of
quark-lepton pairs, and also in GUTs (for example in E$_6$ or Pati-Salam
SO(10) theories). We note that leptoquarks have recently returned to prominence 
in the literature due to the excess of high-$Q^2$ events in Deep Inelastic
Scattering at HERA by both the ZEUS and H1{\cite {zeus}} Collaborations. 
Leptoquarks have fixed gauge couplings to the photon, the $W/Z$  bosons,
and gluons (for spin--1 leptoquarks an anomalous chromo-magnetic moment may
be present), and also {\it a priori} undetermined Yukawa couplings to 
lepton--quark pairs which determine their decays. For phenomenologically 
relevant leptoquarks, this Yukawa coupling should be chiral in 
order to, \eg,  restrain leptons from acquiring too large a 
magnetic moment and to 
prevent large violations in universality from $\pi$ decay. In addition, they
should essentially couple only to a single SM family to 
avoid problems with Flavor Changing Neutral Currents.

The interactions of leptoquarks can be described by an effective low-energy
Lagrangian.  The most general renormalizable $SU(3)_C\times SU(2)_L\times 
U(1)_Y$ 
invariant leptoquark-fermion interactions can be classified by their fermion
number, $F=3B+L$, and take the form\cite{brw}
\begin{equation}
{\cal L}  =  {\cal L}_{F=-2} + {\cal L}_{F=0} \,,
\end{equation}
with
\begin{eqnarray}
{\cal L}_{F=-2} & = & (g_{1L}\bar q^c_Li\tau_2\ell_L+g_{1R}\bar u^c_Re_R)S_1
+\tilde g_{1R}\bar d^c_Re_R\tilde S_1 \nonumber\\
& & +g_{3L}\bar q^c_Li\tau_2\vec\tau\ell_L\vec S_3 \nonumber\\
& & +(g_{2L}\bar d^c_R\gamma_\mu\ell_L+g_{2R}\bar q^c_L\gamma^\mu e_R)V_{2\mu}
\nonumber\\
& & +\tilde g_{2L}\bar u^c_R\gamma^\mu\ell_L\tilde V_{2\mu} + h.c. \,,\\
{\cal L}_{F=0} & = & (h_{2L}\bar u_R\ell_L+h_{2R}\bar q_Li\tau_2e_R)R_2
+\tilde h_{2L}\bar d_R\ell_L\tilde R_2\nonumber\\
& & +(h_{1L}\bar q_L\gamma^\mu\ell_L
+h_{1R}\bar d_R\gamma^\mu e_R)U_{1\mu} \nonumber\\
& & +\tilde h_{1R}\bar u_R\gamma^\mu e_R\tilde U_{1\mu}
+h_{3L}\bar q_L\vec\tau\gamma^\mu\ell_L\vec U_{3\mu} + h.c.
\nonumber
\end{eqnarray}
Here, $q_L$ and $\ell_L$ denote the $SU(2)_L$ quark and lepton doublets,
respectively, while $u_R\,, d_R$ and $e_R$ are the corresponding singlets.
The indices of the leptoquark fields indicate the dimension of their $SU(2)_L$
representation.  The subscripts of the coupling constants label the lepton's 
chirality. For an overview of the phenomenology associated with a 
light, HERA-inspired leptoquark see for example {\cite {lqtj}} and 
references therein.

$ii)$ Diquarks: These particles have B$=\pm 2/3$ and L$=0$, and 
are also predicted in 
composite models as bound states of quark pairs, and in Grand Unified models
(\eg, the model based on the $E_6$ symmetry group). 

$iii)$ Bileptons:  These particles have B$=0$ and L$=0,\pm2$. They 
occur in, \eg, theories where the electroweak gauge group for leptons is 
extended from $SU(2)_L\times U(1)_Y$ to $SU(3)$ and baryon and lepton numbers 
are conserved. They may carry either 0 or 2 units of lepton number and no 
baryon number. Bileptons 
can appear both as scalar and as vector gauge particles 
and can be singly or doubly charged; for instance, doubly charged dilepton 
gauge bosons appear  in $SU(15)$ GUTs.  Bileptons have 
couplings to ordinary gauge bosons which are  fixed by gauge invariance, and 
Yukawa couplings to leptons which mediate their decays. For a detailed survey, 
see Cuypers and Davidson{\cite {cd}}.

Clearly the possible set of new particles is so large that we cannot hope to 
examine production signatures and search reaches for all of the above at 
future hadron and lepton colliders and so we will concentrate on the new work 
that was performed at Snowmass on just a few 
of these possibilities: leptoquark production at lepton and hadron colliders, 
bilepton 
production at the NLC, and neutral heavy lepton production at lepton 
colliders. For a summary of older work on this subject, see Ref. {\cite {dpf}}.

\subsection{New Particle Production at Colliders}

At hadron colliders the best way to search for leptoquarks is through the pair 
production process $q\bar q,gg \to LQ \overline {LQ}$ with the on-shell 
leptoquarks then decaying 
into (1) two jets plus charged leptons, (2) two jets, one charged lepton and 
missing energy or (3) two jets plus missing energy. Clearly the SM backgrounds 
increase as we go from (1) to (3) making discovery difficult. In most 
analyses, leptoquarks 
are considered to be produced `one at a time', \ie, the fact 
that they may lie in nearly degenerate multiplets is usually ignored. 
Fortunately, all leptoquark 
multiplets lead to a rather high branching fraction, 
$B\geq 0.5$, into the charged lepton mode as can be observed by an examination 
of the Lagrangian above. For scalar, \ie, spin-0 leptoquarks the cross section 
depends solely on their mass in the limit that the $q \ell LQ$ Yukawa 
coupling, $\lambda$, is of electroweak strength or less, \ie, 
$\tilde \lambda=\lambda/e <1$. In the vector(spin-1) case the situation is 
somewhat less clear. If vector leptoquarks 
are gauge particles then the cross section depends solely on their mass. 
However, it possible that the cross section may depend on one or 
more additional parameters such at the anomalous chromomagnetic moment, 
$\kappa$. Note that for the case where vector leptoquarks are 
gauge particles $\kappa$ is fixed to unity. 

The production of both scalar and vector leptoquarks 
at the Tevatron and LHC have 
been previously discussed in the literature. Rizzo
updated{\cite {rizzo}}  these analyses and extended them to TeV33 
and possible higher energy hadron colliders at $\sqrt s=60,200$ TeV. 
As is well-known, hadron colliders can distinguish scalar from vector 
leptoquarks from the size of the cross section, and 
perhaps tell us something about their charged lepton branching fraction as well.
However, we cannot learn about the leptoquark's 
electroweak interactions at a hadron collider. For leptonic branching fractions
of unity and a conservative 
assumption about the number of required signal events the search reaches for 
scalar leptoquarks
were found to be 0.35(1.34, 4.9, 15.4)TeV at TeV33, LHC and the two 
higher energy colliders, respectively,  assuming that they ran in $pp$ mode. 
The corresponding reaches for gauge boson vector leptoquarks was found to be 
0.58(2.1, 7.6, 24.2)TeV.  At the LHC a detailed study 
of second generation scalar leptoquark pair production was 
performed using the CMS detector fast Monte Carlo{\cite {wrochna}} in order 
to understand backgrounds and finite resolution effects.  The results are
shown in Fig. 7, where we see that the search
reach may be as high as $M_{LQ}=1.6$ TeV.

\begin{figure}[t]
\leavevmode
\centerline{\epsfig{file=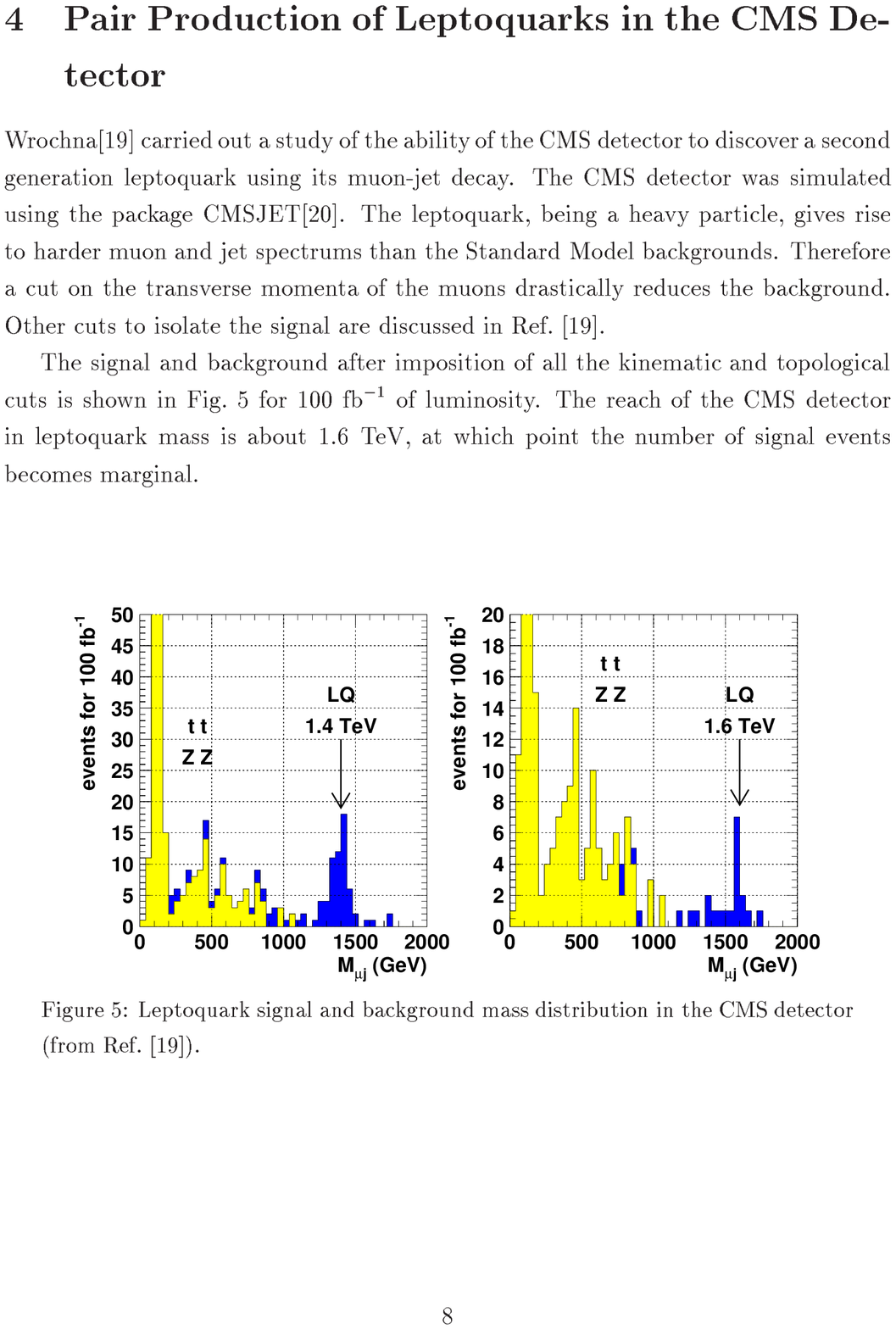,width=9.0cm,clip=}}
\caption{Leptoquark signal and background mass distribution for second
generation leptoquarks for the CMS detector.  (From Ref. \cite{wrochna}.)}
\end{figure}

\begin{figure}[htbp]
\centerline{
\epsfig{figure=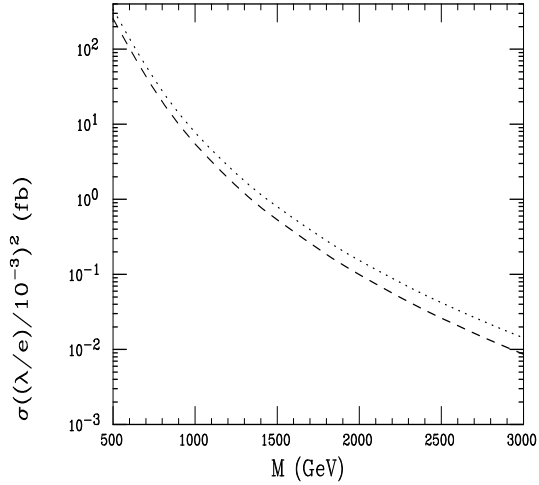,height=8cm,width=8cm,angle=-90}}
\vspace*{-0.9cm}
\caption{Single scalar leptoquark 
production cross sections at a 100 TeV $pp$ collider 
as functions of the leptoquark 
mass, for both the $gu$ (dotted) and $gd$ (dashed) 
intial states. The overall Yukawa coupling has been rescaled to units of 
$\tilde \lambda/10^{-3}$.}
\label{singlelq}
\end{figure}

Another possibility which is not often discussed is the single production of 
leptoquarks
vis $gq$ fusion, \ie, $gq \to LQ +\ell$ where $\ell$ is either a charged 
lepton or a neutrino. The cross section for this depends quadratically on 
the unknown Yukawa coupling $\tilde \lambda$. For sizeable values of 
$\tilde \lambda$ this process will dominate pair production. 
For very small values of $\tilde \lambda$ it 
is clear that the pair production cross section is far larger even though a 
pair of heavy objects is being produced. However, the single production 
process allows one to study the size of the Yukawa coupling for a leptoquark
which has 
already been observed through the pair production mechanism. For example, 
Fig.~\ref{singlelq} shows the single production cross section for a scalar
leptoquark at 
a $\sqrt s=100$ TeV collider for very small values of $\tilde \lambda$. For 
luminosities in the $100-1000 fb^{-1}$ range very large event rates are 
obtained for scalar leptoquarks as heavy as 1.5 TeV.

\vspace*{-0.5cm}
\nn
\vspace*{0.4mm}

Leptoquarks can also be pair produced at lepton colliders. As is well known, 
their 
production characteristics yield complete information about their spin and all 
of their electroweak quantum numbers. The only difficulty is that the pair 
production reach is limited by ${\sqrt s}/2$ and thus much attention has 
focussed on single production of leptoquarks 
via $\gamma e$ collisions through either 
the Weisacker-Williams process or at a true photon-electron collider employing 
the backscattered laser technique. As shown in Fig. 9 from Doncheski and 
Godfrey\cite{steve}, for electromagnetic strength Yukawa couplings the 
search reach is significantly extended in either case and that polarization 
asymmetries can be used to determine the leptoquarks quantum numbers. 
Of course this approach fails if the Yukawa 
couplings are substantially smaller that this assumed 
strength.

\begin{figure}[h]
\leavevmode
\centerline{\epsfig{file=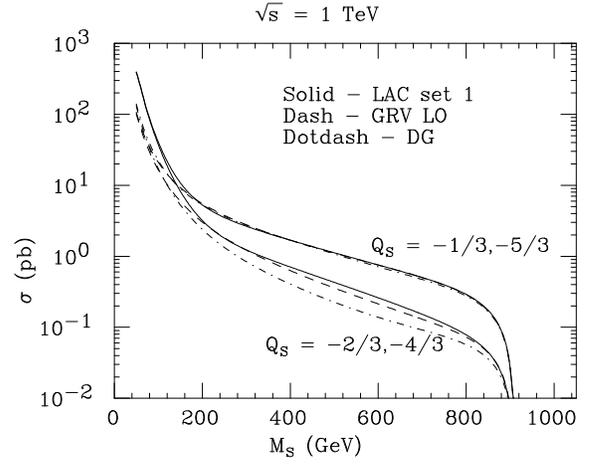,width=9.0cm,clip=}}
\caption{The cross sections for leptoquark production due to resolved 
photon contributions in $e\gamma$ collisions.  The photon beam is due 
to laser backscattering in a $\sqrt{s}=1000$~GeV collider. The 
different curves correspond to different photon distribution functions
(from Ref. \cite{steve}.)}
\end{figure}

If very heavy leptoquarks exist then they may be searched for indirectly in 
$\ell^+ \ell^- \to q \bar q$ since they constitute new $t-$ or $u-$channel 
exchanges. Again the potential size of their influence is controlled by the 
size of their Yukawa couplings. By combining angular and 
polarization asymmetries as well as the total cross section it is possible to 
examine which regions of the Yukawa coupling-LQ mass plane would show 
such sensitivity. This case was analyzed in detail by Berger{\cite {mikeb}} for 
the generic scalar leptoquark 
scenario. Assuming either right- or left-handed couplings for 
the scalar leptoquark and electromagnetic coupling 
strength for the Yukawa couplings, both the NLC and 
the NMC were found to be able to probe scalar leptoquark 
masses in the range $1.5-2\sqrt s$ 
assuming canonical luminosities. 

Cuypers and Davidson{\cite {cd}} have performed a comprehensive examination 
of the search reach for bileptons at the NLC in the $\gamma \gamma$, 
$\gamma e$, $e^+e^-$ and $e^-e^-$ collider modes. All of these modes provide 
a reach up to the kinematic limit and can yield detailed information on the 
bilepton quantum numbers and Yukawa coupling structure. Using various modes, 
the reach for bileptons at the NLC with canonical luminosities was found to be 
$m_{BL}\geq 50\lambda_{ee} \sqrt s$ where $\lambda_{ee}$ is the bilepton
coupling to $ee$, as displayed in Fig. 10.

\begin{figure}[h]
\leavevmode
\centerline{\epsfig{file=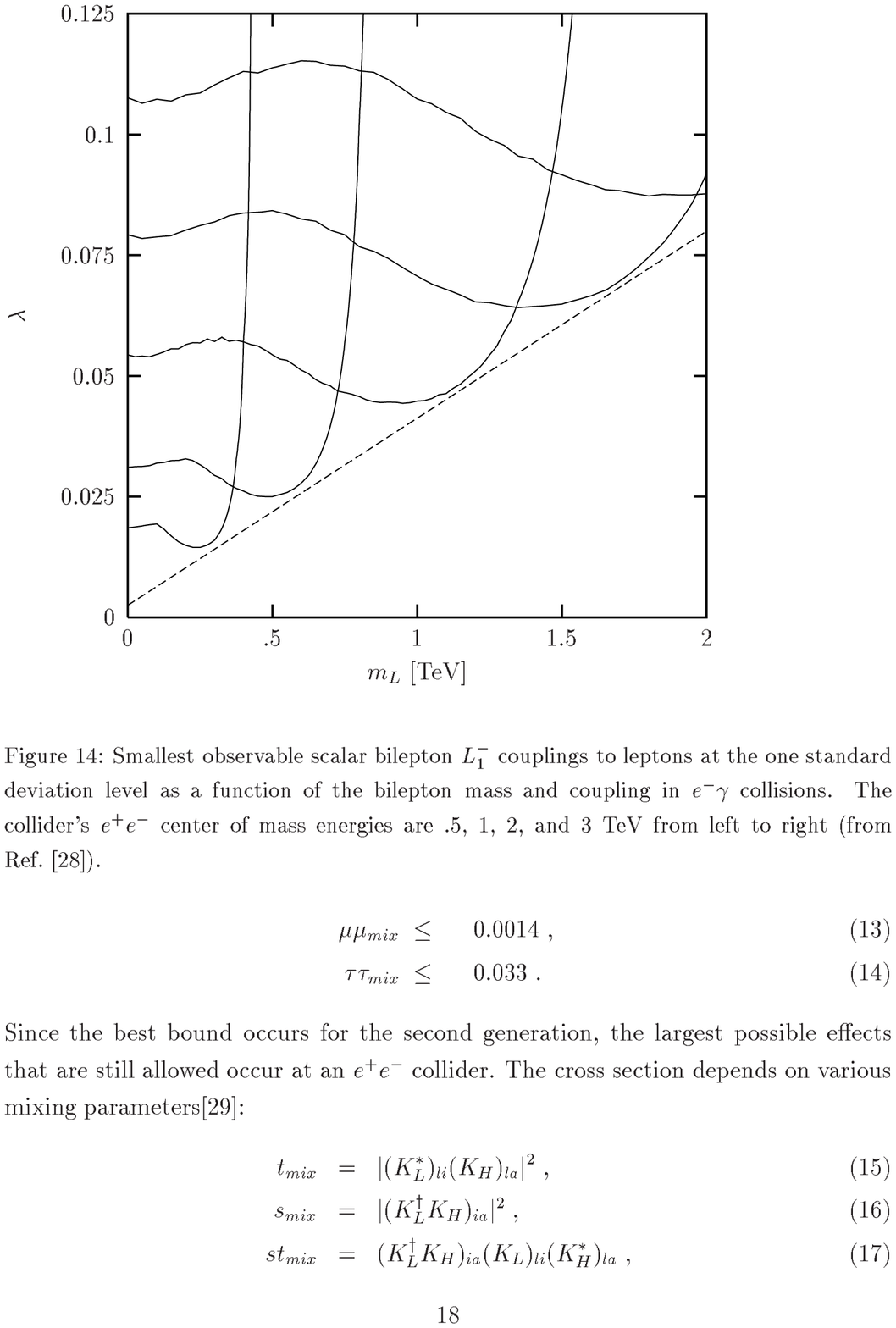,width=6.0cm,clip=}}
\caption{Smallest observable scalar bilepton coupling as a function of the
bilepton mass at the level one standard deviation in $e\gamma$ processes.
The assumed $e^+e^-$ center of mass energies are 0.5, 1, 2, and 3 Tev, from
left to right.  (From Ref. \cite{cd}.)}
\end{figure}

Kalyniak and Melo{\cite {pat}} studied the single production of neutral 
heavy leptons in association with a massless neutrino at lepton 
colliders. These particles may be produced either by $s-$channel $Z$ exchange 
and/or by $W$ exchange in the $t-$channel depending on the leptonic flavor. 
These authors concentrated on a model where every generation has a massless 
neutrino as well as one heavy Dirac neutrino. The cross sections are 
calculated in terms of the heavy lepton masses and a set of mixing parameters 
which describe the experimentally allowed size of the violation of unitarity 
due to mixing amongst all 6 neutrinos in the $3\times 3$ light neutrino basis. 
For mixing parameters as large as allowed by current experiment the search 
reaches were found to extend out to the kinematic limit of a given machine
as shown in Fig. 11.
For masses well inside the kinematic limit, extremely small values of the 
mixing parameters, of order $10^{-(4-6)}$, were found to be accessible.  

\begin{figure}[h]
\leavevmode
\centerline{\epsfig{file=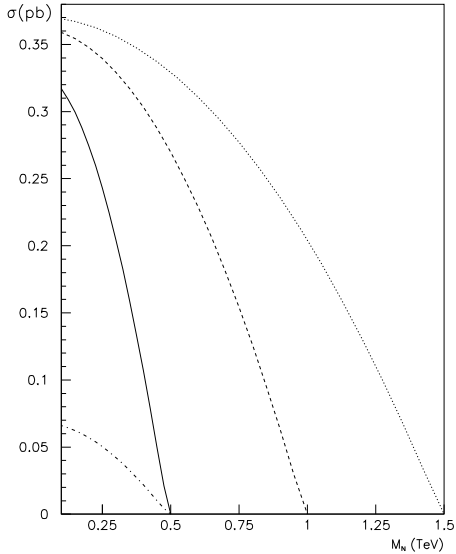,width=6.0cm,clip=}}
\caption{Total cross section vs the neutral heavy lepton 
mass $M_N$ for $e^+e^-$ collider 
at three different energies: $\sqrt{s}=0.5$~TeV (solid line), 
$\sqrt{s}=1.0$~TeV (dashed line), and $\sqrt{s}=1.5$~TeV (dotted line)
and for $\mu^+\mu^-$ collider at $\sqrt{s}=0.5$~TeV (dash-dotted line);
$ee_{mix}=0.0071$, $\mu\mu_{mix}=0.0014$, $\tau\tau_{mix}=0.033$
 (from Ref. \cite{pat}).}
\end{figure}

Heavy neutrinos of the Majorana type are perhaps best probed in $e^-e^-$ 
collisions since the initial state has $L=2$. It was pointed out many years 
ago{\cite {oldtr}} that heavy Majorana neutrinos, exchanged in the 
$t-$ and $u-$channels, might mediate the process 
$e^-e^- \to W^-_{L,R}W^-_{L,R}$ at an observable rate. Since that time there 
has been some controversy concerning whether large rates can be obtained in 
the case where the SM gauge group is not augmented due to constraints from 
other processes, such as the lack of the observation of 
neutrinoless double beta decay. There were two overlapping 
analyses presented at Snowmass on this subject by Heusch{\cite {heusch}} and 
by Greub and Minkowski{\cite {chris}}, who both advocate models where large 
rates may be obtainable for suitable ranges of the parameters. In particular, 
Heusch points out the rather large theoretical uncertainties in the nuclear 
physics aspects of double beta decay in the limit that highly massive objects 
are being exchanged, \ie, in the truly short-distance limit. Heusch argues 
that a number of quark-level inhibition factors arise in this case, which 
when combined reduce the size of the neutrinoless double beta decay 
matrix element by more than a factor of 40. This substantially enlarges the 
parameter space over which $e^-e^- \to W^-_{L}W^-_{L}$ can be sizeable. 
Greub and Minkowski show in a very detailed analysis that the size of the 
resulting cross section can be large and they demonstrate that the backgrounds 
from the more conventional SM processes are small and can be controlled by 
both beam polarization and various kinematic cuts. 

As a summary of new particle production we display in Table II 
the search reaches obtainable at various colliders for the particles surveyed 
here.  We note that where the listed discovery limit is larger than $\sqrt s$, 
we have included the reach from indirect effects.  The question marks in the
Table indicate that a detailed study has not yet been performed.  Figure 12
presents the search reach for new particles decaying into dijets at the
Tevatron, for several possible scenarios\cite{tevstudy}.  
We see here that large integrated luminosities will reach the TeV range.

\begin{figure}[h]
\leavevmode
\centerline{\epsfig{file=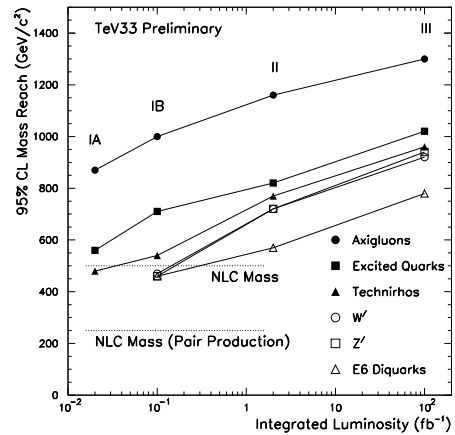,width=6.0cm,clip=}}
\caption{The $95\%$ C.L. mass reach for new particles decaying into dijets as
a function of luminosity at the Tevatron.}
\end{figure}

\begin{table*}[htpb]
\leavevmode
\begin{center}
\caption{New particle discovery reaches in TeV at future colliders. The 
luminosities for TeV33(LHC, 500 GeV NLC, 1 TeV NLC) are assumed to be 
10(100, 50, 100)$fb^{-1}$, respectively. In the case of LQ's at the NLC, the 
first(second) value is for pair production (single production with 
electromagnetic strength Yukawa couplings). The third value is the indirect 
reach in the later case. The question mark indicates that a Monte Carlo
study has not yet been performed.}
\label{reaches}
\begin{tabular}{lcccc}
\hline
\hline
Particle & TeV33     & LHC      & 500 GeV NLC      & 1 TeV NLC   \\
\hline
Scalar LQ      & 0.33      & 1.6      & 0.25,0.45,5.0    & 0.5,0.9,6.5 \\
Vector LQ      &$\sim 0.5$ &$\sim 2.2$ & 0.25,0.45,3.0    & 0.5,0.9,5.5 \\ 
Axigluon & 1.3      &$\sim 5.0$ &  0.4             & 0.8       \\
Heavy Q  & ?        &   ?       & 0.25             & 0.5       \\
Heavy L  & ?        &   ?       & 0.25             & 0.5       \\
Diquark  &$0.2-0.78$  &$\sim 5$   & 0.25             & 0.5       \\
Bilepton & -        &  -        & 0.45             & 0.9       \\
\hline
\hline
\end{tabular}
\end{center}
\end{table*}

\section{New Interactions}

Without knowing what physics lies beyond the standard model we can 
take several distinct approaches.  In previous sections we explored the 
phenomenology of specific manifestations of new physics.  In this 
section we take a more generic approach; looking for new physics 
via the effect they have on interactions well below their typical 
scales.  First we consider models of 
dynamical symmetry breaking and their ``low energy'' particle 
spectrum.  Quite generally, if a low mass Higgs boson does not exist 
and the weak sector becomes strongly interacting at high energy a 
whole spectrum of states should exist, similar to the low lying 
particle spectrum of QCD. This subject 
was studied in detail by another Snowmass working group \cite{siws}.
However, several members of the new phenomena 
working group also examined the 
phenomenology of specific examples of this scenario; one-family 
technicolor and topcolor assisted technicolor.  
Secondly, we take this progression to its 
conclusion, that new particles are not observed and new physics only 
manifests itself through the existence of effective interactions at 
low energy. For the detailed report see the subgroup summary by Cheung and 
Harris\cite{robking}.

\subsection{One-Family Technicolor}

Eichten and Lane \cite{eichten} described a one-family technicolor 
model with color triplet techniquarks and color singlet 
technileptons.  The techniquarks bind to form color singlet 
technirhos, $\rho_{T1}^\pm$ and $\rho_{T1}^0$, with mass roughly in the 
range 200 to 400 GeV.  Color singlet technirhos can be produced in 
hadron collisions through quark-antiquark annihilation.  The expected 
decay modes are $\rho_{T1}^\pm \to W^\pm Z$, $W^\pm \pi_T^0$, 
$Z \pi_T^\pm$, $\pi^\pm_T \pi^0_T$ and $\rho_{T1}^0 \to W^\pm W^\mp$,
$W^\pm \pi^\mp_T$, $\pi^\pm_T \pi^\mp_T$.  The technipions, $\pi_T$, 
in turn decay predominantly to heavy flavors: $\pi^0_T \to b\bar{b}$, 
and $\pi^\pm_T \to c\bar{b}, \; t\bar{b}$.  Techniquarks will also bind 
to form color octet technirhos, $\rho_{T8}^0$ with mass roughly in 
the range 200 to 600 GeV.  Color octet technirhos are produced and 
decay via strong interactions.  If the mass of the colored 
technipions is greater than half the mass of the technirho, then the 
$\rho_{T8}^0$  will decay predominantly to dijets: $\rho_{T8} \to 
gg$.  If colored technipions are light the  $\rho_{T8}^0$
decays to pairs of either color triplet technipions (leptoquarks) or 
color octet technipions.  

\subsubsection{$\rho_{T1} \to W + dijets$ at the Tevatron}

The search for $\rho_{T1}\to WX$, where $X= W,\; Z$, or $\pi_T$, is 
sufficiently similar to the search for a massive $W'$ decaying to $WZ$ 
that Toback \cite{toback2} was able to extrapolate the $W'$ search to higher 
luminosities as an estimate of the search limits for color singlet 
technirhos at the Tevatron.  He considered the decay chain $\rho_T \to 
WX \to e\nu +dijets$, and required both the electron and neutrino to 
have more than 30~GeV of transverse energy, $E_T$.  He required at 
least two jets in the event, one with $E_T > 50$~GeV and the other with 
$E_T>20$~GeV.  The resulting $W+dijet$ mass distribution from 
100~pb$^{-1}$ of CDF data was in good agreement with standard model 
predictions and was used to determine the 95\% C.L. upper limit on the 
$\rho_{T1}$ cross section, shown in Fig. 13.  The acceptance for the 
technirho was assumed to be roughly the same as for a $W'$.  The 
extrapolation to higher luminosities shows that TeV33 (30 fb$^{-1}$) 
should be able to exclude a color singlet technirho decaying to 
$W+dijets$ up to roughly 400~GeV at 95\% C.L..  This covers the 
expected range in the one family technicolor model. 

\begin{figure}[h]
\leavevmode
\centerline{\epsfig{file=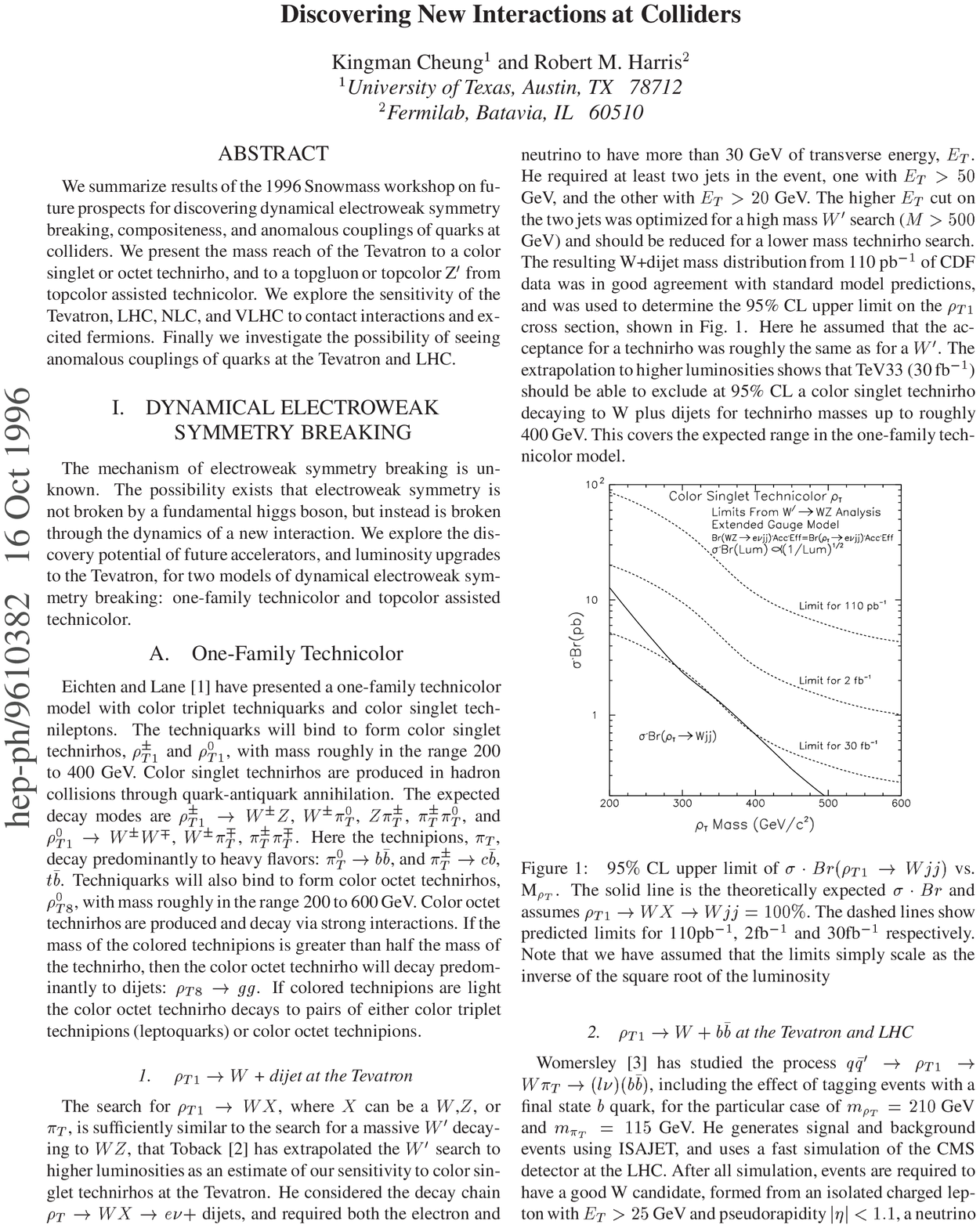,width=6.0cm,clip=}}
\caption{95\% CL upper limit of $\sigma\cdot Br(\rho_T \to Wjj)$ vs 
$M_{\rho_T}$.  The solid line is the theoretically expected 
$\sigma\cdot Br$ and assumes $\rho_{T1} \to WX \to Wjj=100$\%. The 
dashed lines show predicted limits for 100~pb$^{-1}$, 2~fb$^{-1}$, and 
30~fb$^{-1}$ respectively.  (From Ref. \cite{toback2}.)}
\end{figure}

\subsubsection{$\rho_{T1} \to W + b\bar{b}$ at the Tevatron and LHC}

Womersley \cite{womersley} studied the process $q\bar{q}' \to 
\rho_{T1} \to W\pi_T \to (l\nu)(b\bar{b})$, including the effect of 
tagging events with a final state $b$ quark, for the particular case 
of $M_{\rho_T}=210$~GeV and $M_{\pi_T}=115$~GeV.  The signal is a $W$
(reconstructed from $l+\not{\!E}_T$) and two jets with a resonance in 
the dijet mass $m_{jj}$. The backgrounds are $W+jets$ and $t\bar{t}$.
Womersley generated signal 
and background events using ISAJET and used a fast simulation of the 
CMS detector at the LHC.  Events are required to have a good $W$ 
candidate in $l\nu$ mode and two jets with $E_T> 20$~GeV and 
$|\eta|<2.5$.  The single b-tagging efficiency was
assumed to be 50\% with a mistag rate of 1\% for light 
quark jets.  Fig. 14 shows the reconstructed $\pi_T$ peak in the signal 
sample, and that prior to b-tagging the signal is swamped by the large QCD 
$W+jj$ background.  Fig. 14 also shows that after b-tagging the signal 
to background is significantly improved at both the Tevatron and the 
LHC.  The signal to background is better at the Tevatron but the rate 
at the LHC is considerably higher.  Clearly, b-tagging is important to 
reduce the $W+jets$ background.

\begin{figure}[h]
\leavevmode
\begin{minipage}[t]{4.0cm}
\centerline{\epsfig{file=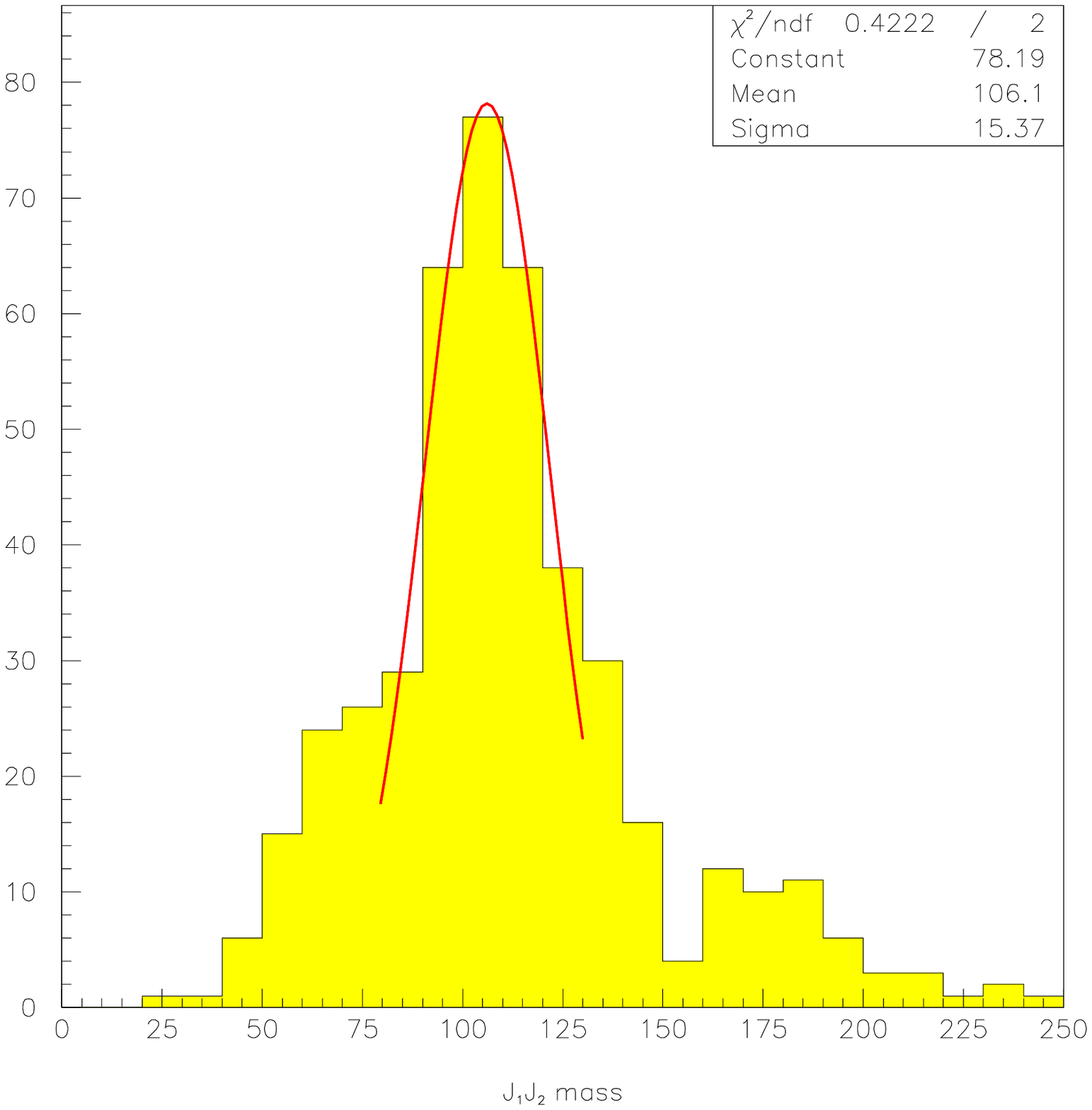,width=4.0cm,
bbllx=37,bblly=150,bburx=535,bbury=654,clip=}}
\end{minipage} \  
\begin{minipage}[t]{4.0cm}
\centerline{\epsfig{file=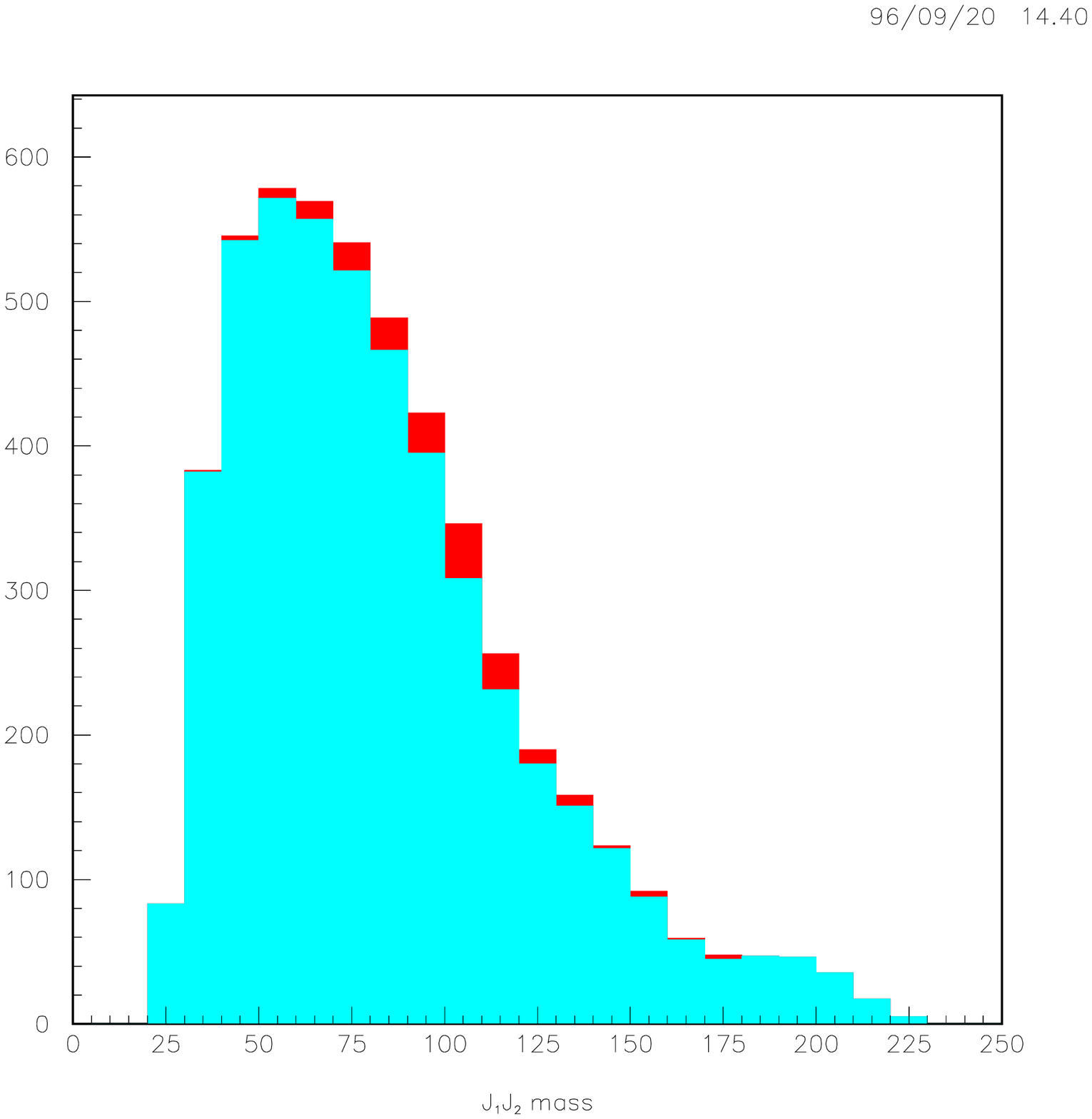,width=4.0cm,
bbllx=37,bblly=150,bburx=535,bbury=654,clip=}}
\end{minipage}
\begin{minipage}[t]{4.0cm}
\centerline{\epsfig{file=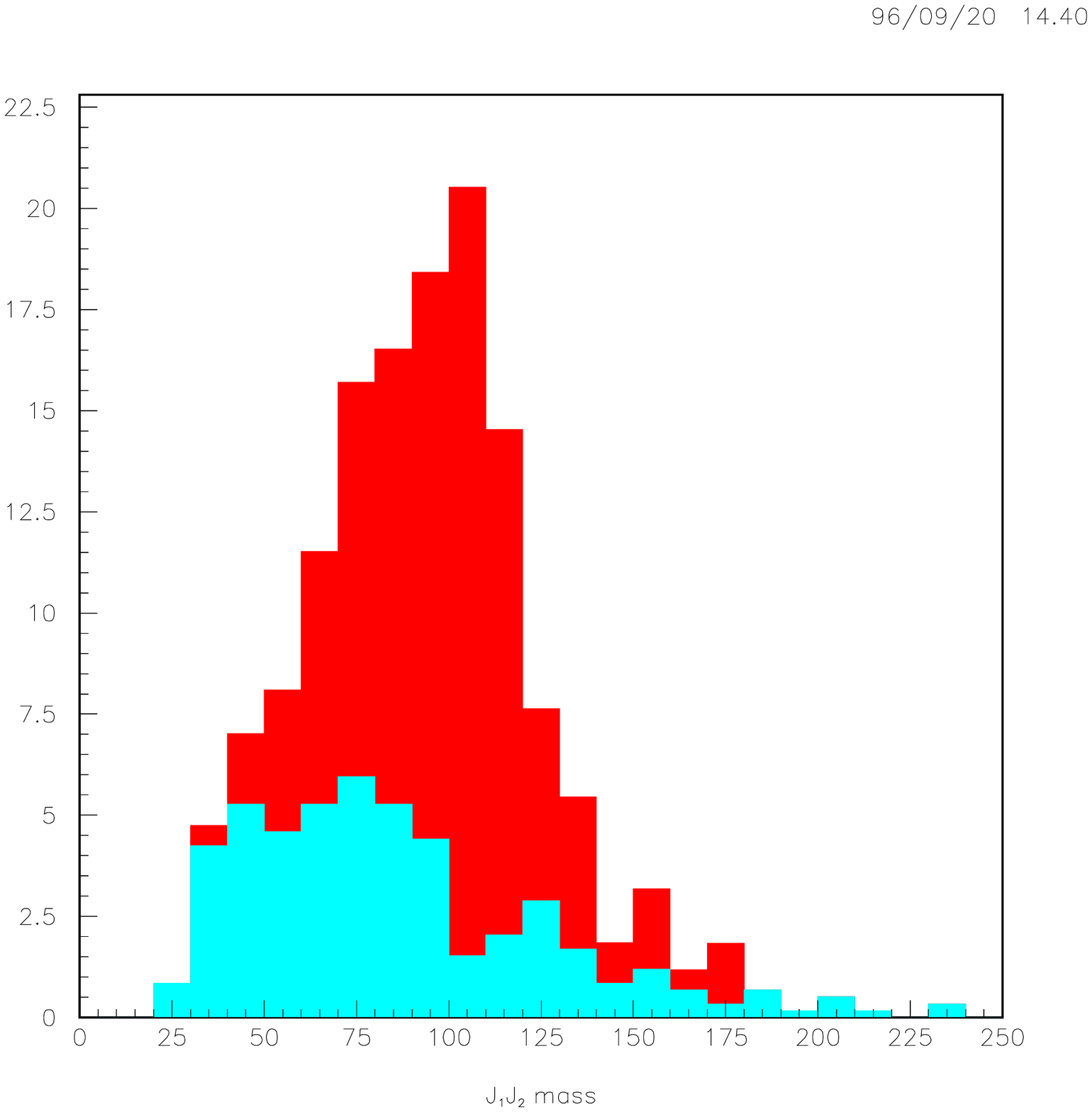,width=4.0cm,
bbllx=37,bblly=150,bburx=535,bbury=654,clip=}}
\end{minipage} \ $\;\;\;\;\;$ \
\begin{minipage}[t]{4.0cm}
\centerline{\epsfig{file=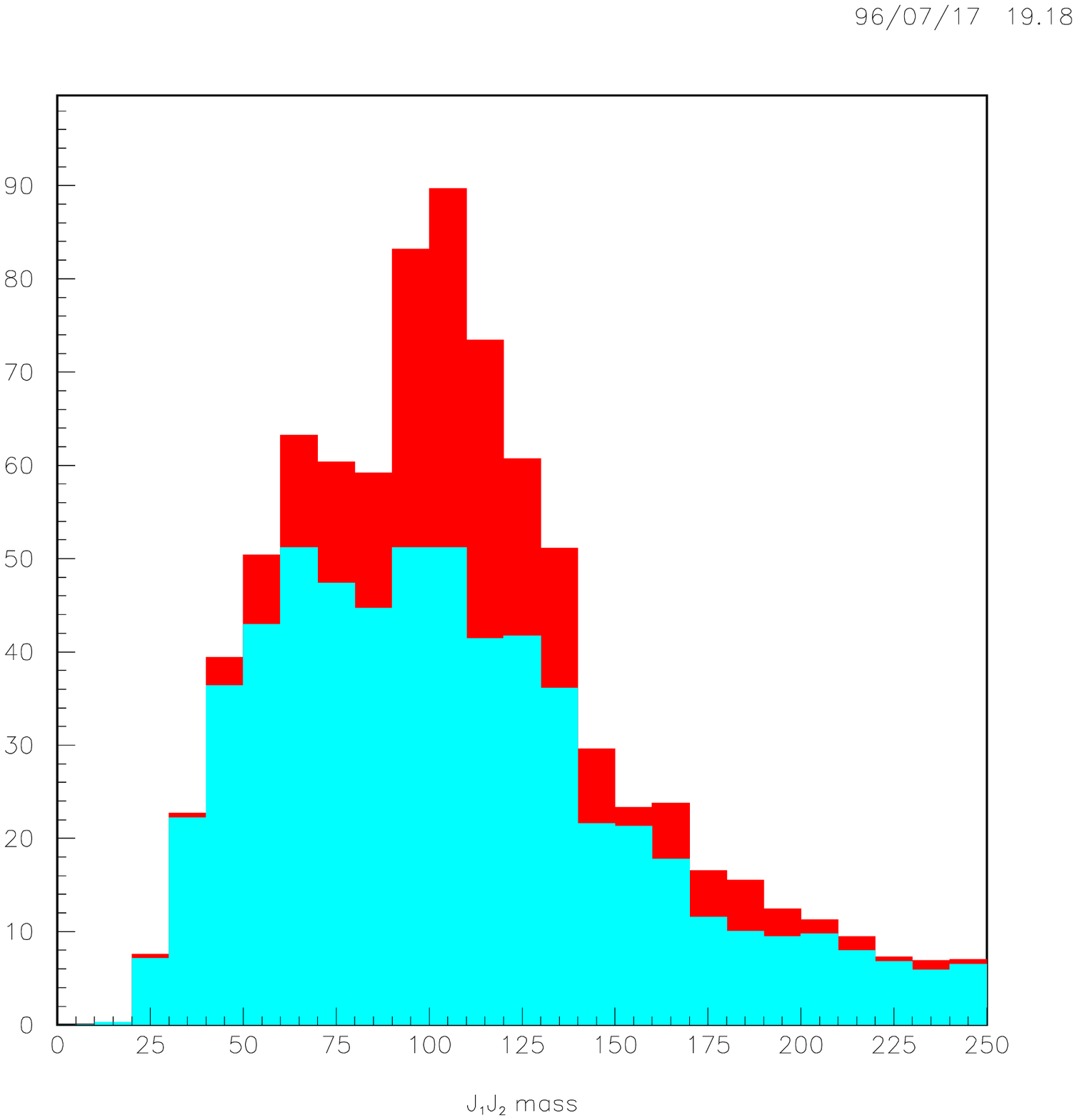,width=4.0cm,
bbllx=37,bblly=150,bburx=535,bbury=654,clip=}}
\end{minipage}
\caption{$\rho_{T1} \to W + \pi_T \to (l\nu)(b\bar{b})$ search.  
(upper left) Leading dijet invariant mass distribution for signal at 
the LHC. (upper right) Same for signal (dark) and background (light) 
at the Tevatron before b-tagging. Vertical scale is events/10 
GeV/2 fb$^{-1}$. (lower left) Same at the Tevatron after b-tagging. 
(lower right) Same at the LHC after b-tagging.  Vertical scale is events/10 
GeV/0.5 fb$^{-1}$. All horizontal scales are in GeV. 
(From Ref. \cite{womersley}.)}
\end{figure}

\subsubsection{$gg \to Z_L Z_L, \; W_L W_L$ at the LHC}

Lee \cite{lee} studied the production of longitudinal gauge boson 
pairs via gluon fusion in the one-family technicolor at the LHC.  
Fig. 15 shows that when the invariant mass is above the threshold for 
production of pairs of colored technipions, the $W_L W_L$ or $Z_L 
Z_L$ signal cross section is greater than the standard model 
background by over an order of magnitude.  Assuming an integrated 
luminosity of 100~fb$^{-1}$, the $Z_L Z_L$ signal, with over a 
thousand events in the four lepton final state ($e$ and $\mu$), will 
be easily observable. 

\begin{figure}[h]
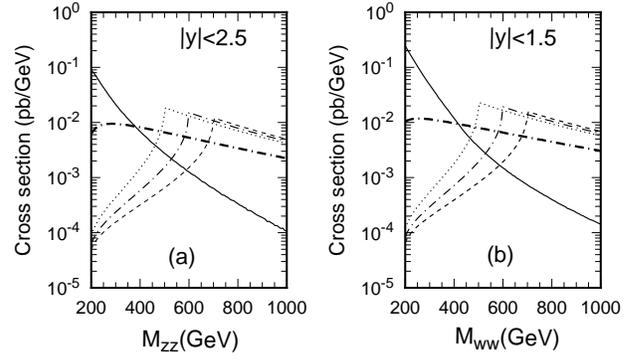

\leavevmode
\begin{minipage}[t]{4.0cm}
\centerline{\epsfig{file=fig4a.ps,width=4.0cm,
bbllx=105,bblly=236,bburx=371,bbury=554,clip=}}
\end{minipage} \
\begin{minipage}[t]{4.0cm}
\centerline{\epsfig{file=fig4b.ps,width=4.0cm,
bbllx=105,bblly=236,bburx=371,bbury=554,clip=}}
\end{minipage}
\caption{The cross sections for (a) $Z_L$-pair and (b) $W_L$ pair 
production via gluon fusion in $pp$ collisions at $\sqrt{s}=14$~TeV.  
The solid curves are for the $q\bar{q}$ initiated backgrounds, and 
dotted, dot-dashed, and dashed curves are for technipion masses of 
250~GeV, 300~GeV, and 350~GeV respectively.  The thick dot-dashed 
curves are for the chiral limit ($m_{\pi_T}=0$).
(From Ref. \cite{lee}.)}
\end{figure}

\subsection{Topcolor Assisted Technicolor}

Theories of dynamical electroweak symmetry breaking, such as extended 
technicolor (ETC), have difficulty in generating large fermion masses, 
particularly $m_t$.  This is circumvented in top quark condensation models, 
where the massive top quark is acquired from $\langle\bar t_Lt_R\rangle$;
however these dynamics alone do not fully break the electroweak symmetry.
The necessary ingredients to accomplish both tasks are present in Topcolor
assisted technicolor\cite{topc}, where electroweak interactions are broken
via technicolor with ETC, and the large top quark mass is obtained from the
combination of ETC and a dynamical condensate.  In this case the new strong 
dynamics is a result of an extended gauge sector which generates the
four fermion interaction
\begin{equation}
{g^2\over\Lambda^2}\bar\psi_Lt_R\bar t_R\psi_L \,,
\end{equation}
where $\psi_L$ is the third generation $SU(2)_L$ quark doublet and 
$\Lambda\simeq {\cal O}(1\tev)$ is the typical scale of the new interactions.
The SM gauge group is thus enlarged to
\begin{equation}
SU(3)_1\times SU(3)_2 \times U(1)_1\times U(1)_2 \to SU(3)_C\times U(1)_Y \,,
\end{equation}
where the $SU(3)_1$ and $U(1)_1$ couple strongly to the third generation.  The
breaking of the $SU(3)$ factors gives rise to a set of massive degenerate
color octet bosons, $B_\mu^a$, with masses $\lsim 2\tev$, as well as the usual
massless gluons.  Here we denote the massive color octet as topgluons, but they
are sometimes referred to as colorons in the literature\cite{simmons}.
Additional interactions, represented in this model by the
extra $U(1)$ factor, must also be present in order to avoid b-quark 
condensation.  (This can also be achieved in axial Topcolor models, where the
$b_R$ field does not couple to the $SU(3)_1$, however this possibility will
not be discussed here.)  The additional $U(1)$ gives rise to a Topcolor $Z'$
boson, which is expected to have mass $\lsim 2-3\tev$.  Constraints on this
$Z'$ from electroweak precision measurements have been considered in Ref.
\cite{chiv}, with the result that $M_{Z'}\gsim 0.5-1.5\tev$ for small values
of the $U(1)$ mixing angle.

Harris\cite{robtop} has examined the production of topgluons, decaying into 
$t\bar t$, as well as the QCD background for this process, and has included
the projected experimental efficiency for reconstructing the $t\bar t$ final
state, in order to estimate the topgluon search reach in a $t\bar t$
resonanant state.  The results are presented in Fig. 16, assuming that the
width of the topgluon is given by $0.3 M$, where $M$ represents the mass of
the topgluon.  We see that the discovery reach probes the Tev scale for
luminosities $> 2$ fb$^{-1}$.

Tollefson\cite{tevstudy} has investigated the discovery reach for topcolor
$Z'$ bosons at the Tevatron, by examining the decay chain $Z'\to t\bar t\to
\ell\nu b\bar bjj$.  Using PYTHIA Monte Carlo, CDF detector simulation, and
full reconstruction for both signal and background, she obtains the results
displayed in Fig. 17a.  Here, a potential 5$\sigma$ resonance signal is 
compared to the expected $Z'$ production cross section.  We see that the search
reach for a narrow $Z'$ approaches the TeV scale at TeV33.  Rizzo 
examined\cite{ngb-sum} the indirect search reach for a topcolor $Z'$ at the NLC.
The results are shown in Fig. 17b, where we see that topcolor $Z'$ bosons with
masses in excess of 4.5 TeV may be discerned from examining charm and top quark
pair final state production.  This explores the entire expected mass region for
the existence of these particles.

\begin{figure}[h]
\leavevmode
\centerline{
\centerline{\epsfig{file=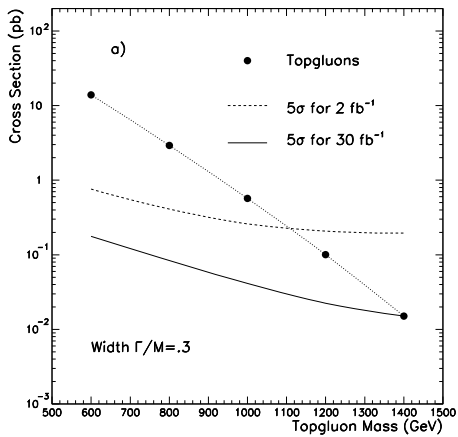,width=6.0cm,
bbllx=17,bblly=15,bburx=534,bbury=525,clip=}}
} 
\caption{ The mass reach for $t\bar t$ decays of topgluons of width $0.3$ M.
The production cross section (points) is compared to the $5\sigma$ reach
at the Tevatron with 2 fb$^{-1}$ (dashed) and 30 fb$^{-1}$ (solid).}
\end{figure}

\begin{figure}[h]
\leavevmode
\centerline{
\centerline{\epsfig{file=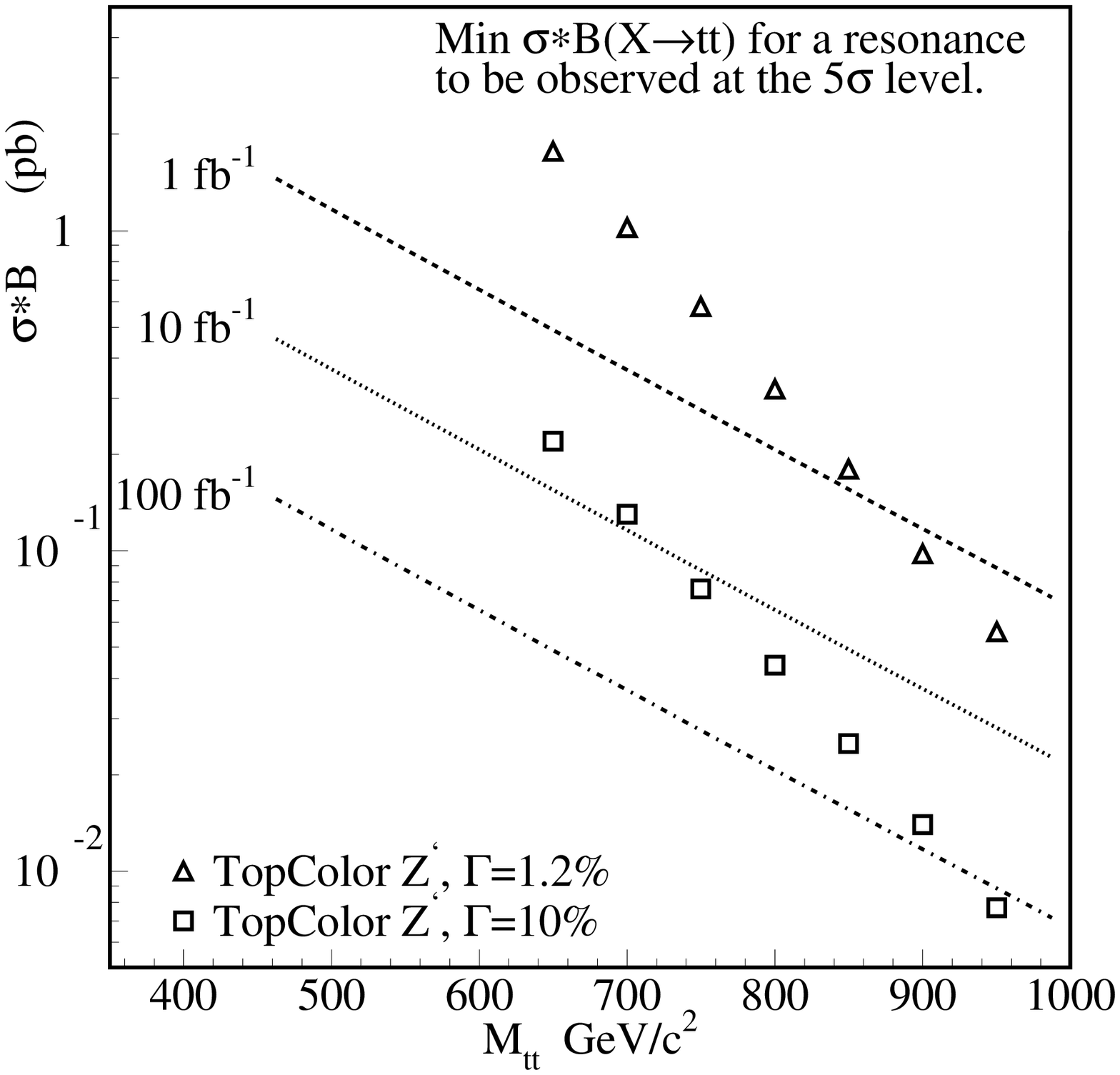,width=5.0cm,
bbllx=6,bblly=3,bburx=535,bbury=518,clip=}}
} 
\centerline{
\begin{turn}{-90}
\epsfig{file=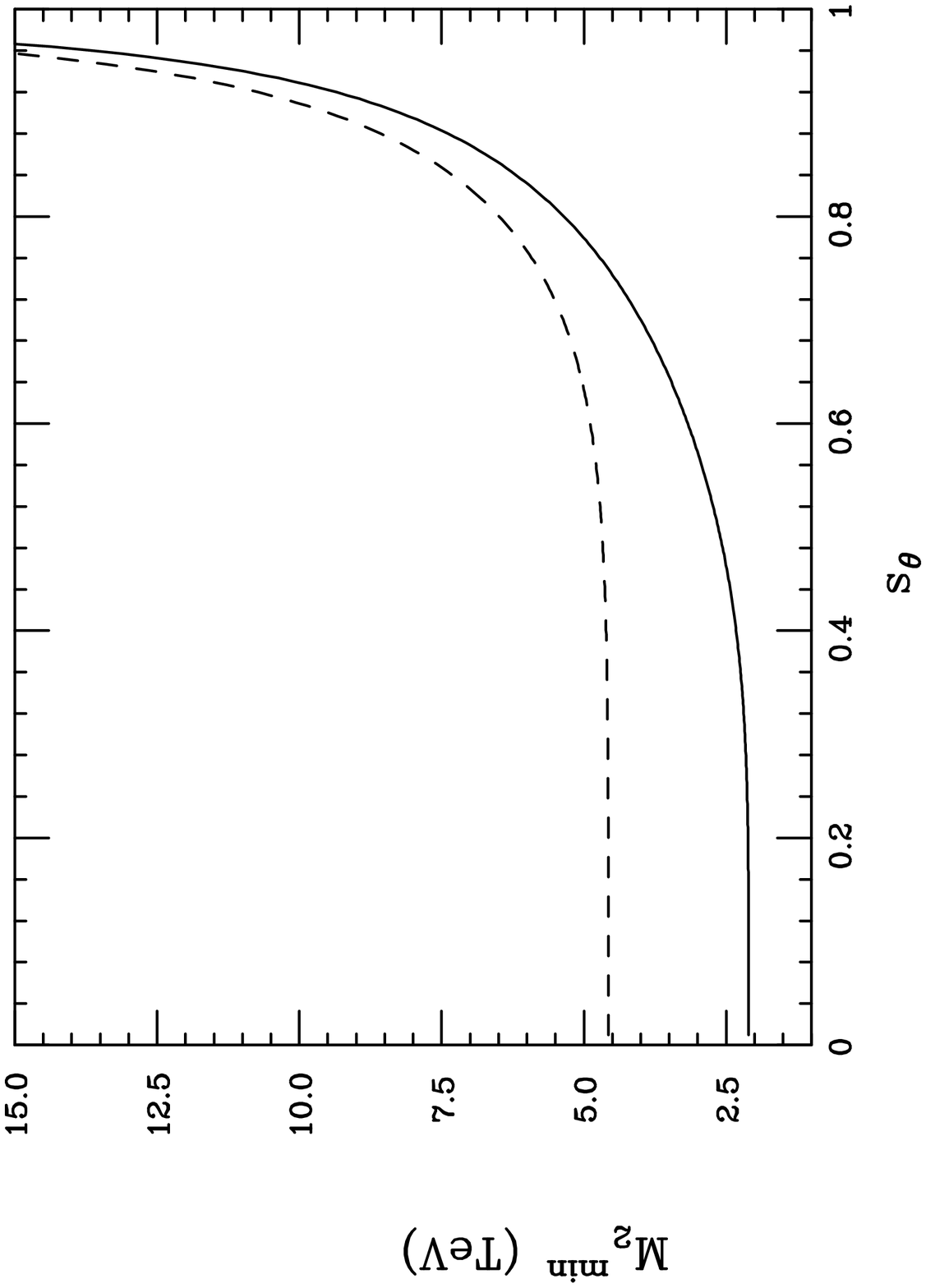,width=5.0cm,clip=}
\end{turn}
} 
\caption{Topcolor $Z'$ search reach at  (a) TeV33 as a function of its mass,
where the width is assumed to be $0.012$M (triangles) or $0.1$M (squares).
(b) a 500 GeV NLC with 50 fb$^{-1}$, where the solid lines include data
from $e, \mu, \tau$ and $b$ final states, and the dashed curve also includes
data on $c$ and $t$.}
\end{figure}

\subsection{Effective Operators}

As we have seen in previous sections, new physics 
can show up either through the direct production of new particles or 
through deviations of precision measurements from their standard model 
predictions.  A powerful and systematic
approach to parametrizing the effects of new 
physics is to use an effective Lagrangian (${\cal{L}}_{eff}$)
with the various terms 
ordered in terms of an energy expansion of the scale of new physics:
\begin{equation}
{\cal{L}}_{eff}=   {\cal{L}}_0 + \frac{1}{\Lambda}{\cal{L}}_1
+ \frac{1}{\Lambda^2}{\cal{L}}_2 + \ldots
\end{equation}
In this approach, an effective Lagrangian obeys the SM 
symmetries and is constructed out of the SM fields.  The leading terms are 
given by the SM while the coefficients of higher dimension operators 
parametrize the effects of new physics.
The effective Lagrangian for an analysis of new interactions was
written down by Buchm\"uller and Wyler \cite{b-w}.  Specific examples 
of new physics will modify the coefficients of ${\cal{L}}_{eff}$ in 
unique patterns characteristic of the new physics.  
For example, Layssac {\it et al.} have shown that the existence of a 
$Z'$ would result in a unique pattern of deviations \cite{layssac}.
Thus, if deviations were observed the pattern would give important 
clues to the underlying physics.  In this section we follow this 
approach to find how precisely future colliders can measure 
several of the terms in ${\cal{L}}_{eff}$.  While we concentrate on specific 
terms, related analysis putting constraints on other types of new 
physics can be found in reports of the Strongly Interacting Weak 
Sector working group.

\subsubsection{Contact Interactions}

Contact, or four Fermi, interactions have long been used to parametrize 
the effects of fermion substructure as form factors or the residual 
effects of constituent binding forces.  Nevertheless one of the first 
manifestations of contact interactions was Fermi decay with its 
characteristic coupling, $G_F$,  having dimensions of inverse mass 
squared and which we now know to be a low energy approximation to 
$W$-boson exchange.  In a previous section we saw how the effects of
new gauge bosons could be observed through deviations of precision 
measurements with contact terms proportional to  $1/M_{Z'}^2$
(for a fixed $\sqrt{s}$).  One can imagine that contact interactions 
can signal many other types of new physics.
So we see that contact interactions can 
indicate many types of physics beyond the standard model, with the 
pattern of new interactions pointing to the nature of the new physics.

In this subsection we consider contact interactions for $(f 
\bar{f})(f'\bar{f}')$. The lowest order four-fermion contact 
interaction, of dimension 6, can be written most generally as
\begin{equation}
{\cal L}  = \frac{g^2_{eff}\eta}{2\Lambda^2} \biggr( \bar q \gamma^\mu q + 
 {\cal F}_\ell \bar \ell \gamma^\mu \ell \biggr )_{L/R}
\; \biggr( \bar q \gamma_\mu q + 
 {\cal F}_\ell \bar \ell \gamma_\mu \ell \biggr )_{L/R} \;,
\end{equation}
where the generation and color indices have been suppressed, $\eta=\pm 1$, and
${\cal F}_\ell$ is inserted to allow for different quark and lepton couplings
but is anticipated to be ${\cal O}(1)$.   Since 
when these operators are used to parametrize substructure, 
the binding force is expected
to be strong when $Q^2$ approaches $\Lambda$, it is conventional to define 
$g^2_{eff}=4\pi$.  However, it should be remembered that if other 
types of new physics give rise to these operators, $g^2$ could be much smaller.
The subscript $L/R$ indicates that
the currents in each parenthesis can be either left- or right-handed and
various possible choices of the chiralities lead to different predictions
for the angular distributions of the reactions where the contact terms
contribute.  
Contact interactions can affect jet production, the Drell-Yan process, 
lepton scattering etc.
Compared to the SM, the contact interaction amplitudes are of order  
$\hat{s}/\alpha_s\Lambda^2$ or $\hat{s}/\alpha_{em}\Lambda^2$
so the effects of the contact interactions will be most important 
at large $\hat{s}$.

\paragraph{$\ell \ell \ell'\ell'$ Contact Terms:}
Cheung, Godfrey, and Hewett \cite{cgh} studied 
$\ell\ell \ell'\ell'$ contact interactions at future 
$e^+e^-$ and $\mu^+\mu^-$ colliders and derived limits on the new 
physics scale $\Lambda$ using the reactions $e^+e^-\to \mu^+\mu^-$ and
$\mu^+\mu^-\to e^+e^-$.  For illustration, we show in Fig. 18, the
$\cos\theta$ distribution 
for $e^+e^-\to b\bar{b}$ at $\sqrt{s}=0.5$~TeV for the SM and with the
contact term present for $\eta=1$ and various values of $\Lambda$.
The effects of the contact term are 
qualitatively similar for other final states.
To obtain the sensitivity to the 
compositeness scale they set $\eta=\pm 1$ and performed a $\chi^2$
analysis, 
comparing the angular distributions for a finite value of $\Lambda$ to 
the SM predictions.  The detector acceptance was taken to be 
$|cos\theta|<0.985$ for the $e^+e^-$ collider and 
$|cos\theta|<0.94$ for the $\mu^+\mu^-$ collider.  The angular 
distribution was divided into 10 equal bins.  
The 95\% CL that can be obtained on $\Lambda$ are 
given in Table III.  The sensitivity to $\Lambda$ in contact 
interactions was found to range from 10 to 50 times the center of mass 
energy.  Polarization in the $e^+e^-$ colliders gives slightly higher 
limits than those obtained at unpolarized $\mu^+\mu^-$ colliders of 
the same energy.

In addition to the $e^+e^-$ mode,
Kumar examined the physics reach of fixed target M\/oller scattering 
at the NLC \cite{kumar}.  
He found that $\Lambda_{ee}$ could be probed to roughly 50~TeV in this 
manner.

\paragraph{$\ell\ell q\bar{q}$ Contact Terms:}

Cheung, Godfrey, and Hewett \cite{cgh} also considered the $\ell\ell 
q\bar{q}$ contact interactions at future $e^+e^-$ and $\mu^+\mu^-$ 
colliders.  They restricted themselves to $\ell\ell c\bar{c}$ and 
$\ell\ell b\bar{b}$ terms where the heavy flavor final states can be 
tagged.  They used the same $\chi^2$ analysis described above and 
assumed flavor tagging efficiencies of $\epsilon_b=60\%$ and
$\epsilon_c=35\%$ ($\epsilon_b=60\%$ and $\epsilon_c=35\%$) for the 
$e^+e^-$ ($\mu^+\mu^-$) colliders.  They found that using polarized 
$e^-$ beams could probe slightly higher mass scales than the 
$\mu^+\mu^-$ case, and were potentially very important for 
disentangling the chiral structure of contact terms if they were 
observed.  More importantly, the higher tagging efficiencies at 
$e^+e^-$ colliders results in higher limits for $\Lambda(\ell\ell q\bar{q})$ 
than can be obtained at $\mu^+\mu^-$ colliders (up to a factor of two 
for the $c\bar{c}$ case). 
These results are included in Table III.

\begin{figure}[h]
\leavevmode
\centerline{\epsfig{file=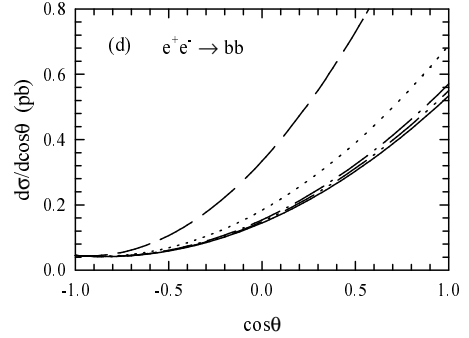,width=6.0cm,clip=}}
\caption{The $\cos\theta$ distribution for unpolarized
$e^+e^- \to b\bar{b}$ at $E_{CM} =0.5$ TeV.
The solid line is for the SM ($\Lambda= \infty $).
Unpolarized $e^+e^-$ with $\eta_{LL}=+1$. 
Dashed line for $\Lambda= 5 $~TeV,
dotted line for $\Lambda= 10 $~TeV,
dot-dashed for $\Lambda= 20 $~TeV, 
and dot-dot-dashed for $\Lambda= 30 $~TeV.}
\end{figure}

P. de Barbaro {\it et al} \cite{barbaro} studied the effect of a left 
handed contact interaction between quarks and leptons at the 
Tevatron.  They compared the invariant mass distribution of 
$\ell\bar{\ell}$ pairs produced in Drell-Yan production for the SM 
with that obtained assuming a left-handed $q\bar{q}\ell\bar{\ell}$contact
interaction for various values of the scale, $\Lambda$. 
A contact interaction would result in an enhancement of the dilepton 
differential cross section at high invariant mass. Fig. 19 shows the 
Drell-Yan cross section for various values of the scale 
$\Lambda_{LL}^-(ee)$.
Barbaro {\it et al} estimated the sensitivity 
of Tevatron measurements with higher luminosities using Monte Carlo 
simulations.
Using 110~pb$^{-1}$ of CDF data on dielectron production 
they report preliminary limits of 
$\Lambda_{LL}^-(q\bar{q}\to e^+e^-)\geq 3.4$~TeV and
$\Lambda_{LL}^+(q\bar{q}\to e^+e^-)\geq 2.4$~TeV at 95\% CL.  Combined 
with the dimuon channel they obtain limits about 0.5 TeV more 
stringent than with electrons alone.    Assuming present detector 
performance, with 30~fb$^{-1}$ of integrated luminosity for TeV33 the
Tevatron will be sensitive to $\Lambda^+_{LL}\leq 14$~TeV and 
$\Lambda^-_{LL}\leq 20$ in the $ee$ channel.

\begin{figure}[h]
\leavevmode
\centerline{\epsfig{file=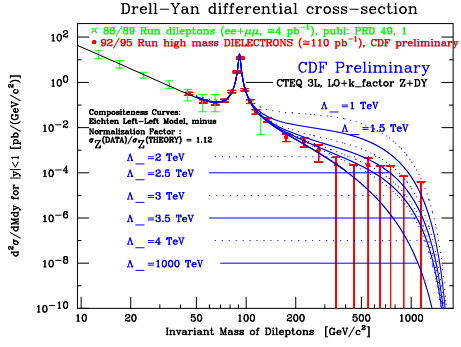,width=6.0cm,clip=}}
\caption{Comparison between CDF Drell-Yan cross-section measurement 
and the theoretical predictions for various values of the scale 
$\Lambda_{LL}(qqee)$ (for $\eta=-1$) for the dielectron channel.
From ref. \cite{barbaro}}
\end{figure}

\paragraph{$ q\bar{q} \to q\bar{q}$ Contact Terms:}

An excess of events with high $E_T$ jets in hadron collisions is a 
well known signature for $ q\bar{q} \to q\bar{q}$ contact interactions.
However the uncertainties in the parton distributions, ambiguities in 
QCD calculations, and uncertainties in jet energy measurement make it 
difficult  to discover a $qq\to qq$ contact interaction.  Another 
signal of $qq\to qq$ contact interactions which is not very sensitive 
to these problems is a dijet angular distribution which is more 
isotropic than predicted by QCD.  Using this approach gives the limits 
on new physics scales given in Table II \cite{abe,atlas}.

\paragraph{$ q\bar{q} \to \gamma\gamma$ Contact Terms:}

The lowest dimension gauge invariant operator involving two fermions 
and two photons is a dimension 8 operator which induces a $q\bar{q} 
\gamma\gamma$ contact interaction.  Assuming parity and CP 
conservation, this interaction is given by:
\begin{equation}
{\cal L} = \frac{2ie^2}{\Lambda^4} Q_q^2 F^{\mu\sigma} F^\nu_\sigma 
\bar{q} \gamma_\mu \partial_\nu q
\end{equation}
where $e$ is the electromagnetic coupling, and $\Lambda$ is the 
associated mass scale.  The observation of the signatures  associated 
with this operator would be a clear signal of compositeness.  
Rizzo analyzed the effects of a $ q\bar{q} \to \gamma\gamma$ contact 
interaction at hadron colliders \cite{rizzo-ci}.
Fig. 20 shows the integrated event rates for 
isolated diphoton events with invariant mass larger than 
$M_{\gamma\gamma}^{min}$ at the LHC  with 100~fb$^{-1}$ 
luminosity.  The cross section is changed most at high 
$M_{\gamma\gamma}^{min}$.  In addition the photon pair will tend to be 
more central with higher average values of $p_T$.
The limits that can be obtained at various 
hadron colliders are given in Table III. The results show that for a 
given center of mass energy the $p\bar{p}$ colliders probe higher mass 
scales than $pp$ colliders because of the higher $q\bar{q}$ 
luminosity.  

\begin{figure}[h]
\leavevmode
\centerline{
\begin{turn}{90}
\epsfig{file=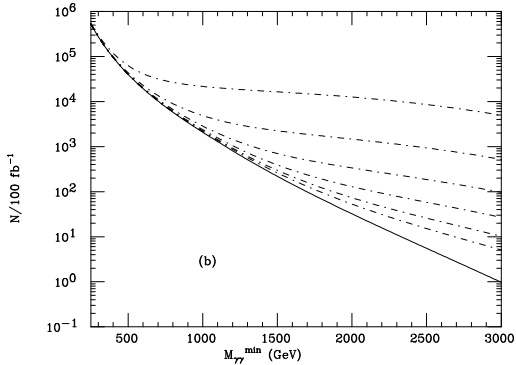,width=6.0cm,clip=}
\end{turn}
} 
\caption{Diphoton pair event rate for the LHC with 100~fb$^{-1}$ as a 
function of $M_{\gamma\gamma}^{min}$
subject to the cuts $p_t^\gamma >200$~GeV and $|\eta_\gamma|<1$.  
From top to bottom the dash-dot curves correspond to $\Lambda_+=0.75$,
1.0, 1.25, 1.75, and 2.0 TeV.  The solid curve is the QCD prediction.}
\end{figure}

\subsubsection{Anomalous Couplings}

The lowest order interactions between a quark and gauge boson are 
dimension 4 and 5 operators of the form
\begin{equation}
{\cal L}_{eff} =g_s \bar{q} T^a \left[ { -\gamma^\mu G_\mu^a
+ \frac{\kappa}{4m_q} \sigma^{\mu\nu} G_{\mu\nu}^a 
- \frac{i\tilde{\kappa}}{4m_q} \sigma^{\mu\nu} \gamma^5 G_{\mu\nu}^a 
}\right] q.
\end{equation}
This particular case corresponds to interactions between a quark and 
gluon where $\kappa/2m_q$ and $\tilde{\kappa}/2m_q$ correspond to
chromomagnetic (CMDM) and chromoelectric (CEDM) 
dipole moment couplings of quarks.  
There are analogous expressions for couplings to the $\gamma$ and $Z$.
Although these couplings are zero 
at tree level, within the SM they can be induced at loop level.
A related example would be $b\to s\gamma$.
One should be cautioned that 
although the factors in the denominator are taken by convention to be 
$m_q$ so that these terms may be expressed as quark dipole moments, 
strictly speaking, $\Lambda$, 
the scale characteristic of substructure or other new physics should 
be used.
These dipole moment couplings are 
important because they are only suppressed by one power of $\Lambda$ 
and also because a nonzero value of the CEDM is a clean signal for CP 
violation.  The above Lagrangian is valid for both light and heavy 
quarks.  In addition to describing an effective $qqg$ vertex it also 
induces a $qqgg$ interaction which is absent in the SM.

Cheung and Silverman studied the effects of anomalous CMDM and CEDM of 
light quarks on prompt photon production \cite{cs-1}.  Prompt photon 
production is sensitive to the gluon luminosity inside a hadron 
because it is mainly produced by quark-gluon scattering.  For the same 
reason it is also sensitive to anomalous couplings of quarks to 
gluons.  The contributing subprocesses for prompt photon production are
$q(\bar{q})g \to \gamma q (\bar{q})$ and $q\bar{q} \to \gamma  g$.  
The prompt photon $p_T$ spectrum is shown in Fig. 21 for the SM and 
for nonzero values of $\kappa$.  Nonzero values of $\kappa$ increases 
the cross section in the high $p_T(\gamma)$ region.   They found that 
CDF and D0 data excludes $\kappa' =\kappa/2m_q < 0.0045$~GeV$^{-1}$.  
However, as stated above, if we rescale $\kappa'$ with a value of 
$\Lambda=1$~TeV we find $\kappa <4.5$.  Naively, we would expect 
$\kappa \sim {\cal O}(1)$.  Silverman and Cheung further estimated the 
sensitivity of the Tevatron and LHC to anomalous CMDM's of light 
quarks.  By binning the jet $E_T$ distributions such that each bin 
would have at most a 10\% statistical error from SM QCD they obtained 
the sensitivities to $\kappa' \equiv 1/\Lambda$ given in Table III
\cite{cs-2}.  
Note that these sensitivities are based on $1-\sigma$ or 68\% CL. 
The effects due to nonzero CEDM will be the same because the increase 
in cross section is proportional to $(\kappa^2 +\tilde{\kappa}^2)$. 

\begin{figure}[h]
\leavevmode
\centerline{\epsfig{file=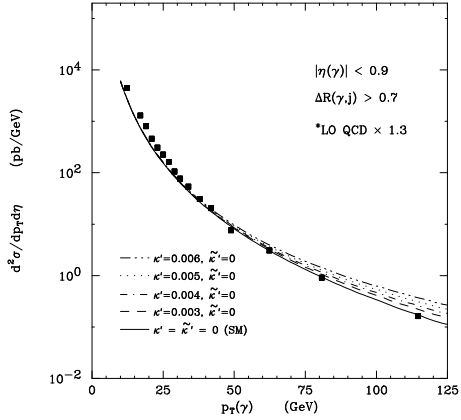,width=6.0cm,clip=}}
\caption{$d^2\sigma/dp_Td\eta$ in prompt $\gamma$ production
for pure QCD and nonzero values of 
CMDM of quarks.  The data points are from CDF. (From Ref. \cite{cs-1}.)}
\end{figure}

Because of the top quark's large mass it is believed by many that the 
detailed physics of the top quark may be significantly different than 
the SM and that the top quark may provide a window into physics beyond 
the SM.  
Rizzo examined anomalous top quark couplings to gluons via top quark 
production at hadron colliders \cite{rizzo-atqc1} and $e^+e^-$ 
colliders \cite{rizzo-atqc2}.  
At hadron colliders the contributing subprocesses to top 
pair production are $q\bar{q},\; gg \to t\bar{t}$.  The existence of a 
nonzero CMDM will change both the total and differential cross 
sections.  The higher center-of-mass energies at the LHC will probe 
beyond the top-pair threshold which will result in much higher 
sensitivities to the CMDM.  $d\sigma/dM_{tt}$ and $d\sigma /dp_T$ 
distributions are shown for the SM and with anomalous couplings in 
Fig. 22 for LHC energies.  
Non-zero $\kappa$ leads to enhanced cross sections at large 
$p_T$ and $M_{tt}$. Rizzo estimated the sensitivities of these distributions 
to anomalous couplings using a Monte Carlo approach and taking into 
account systematic errors.  The 95\% CL for $\kappa$ of the top quark 
are $-0.09 \leq \kappa \leq 0.10$ and $|\kappa| \leq 0.06$ for the 
$M_{tt}$ and $p_t$ distributions respectively.

\begin{figure}[h]
\leavevmode
\centerline{
\begin{turn}{-90}
\epsfig{file=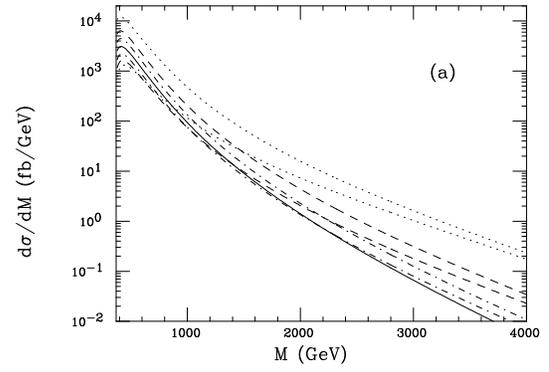,width=6.0cm,clip=}
\end{turn}
} 
\centerline{
\begin{turn}{-90}
\epsfig{file=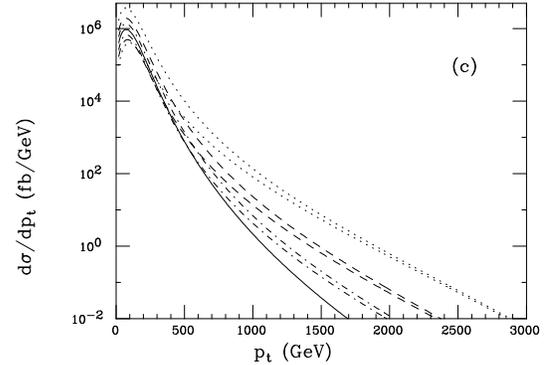,width=6.0cm,clip=}
\end{turn}
} 
\caption{(a) $t\bar{t}$ invariant mass distribution at the LHC for 
various values of $\kappa$ assuming $m_t=180$~GeV. (c) $t\bar{t}$ 
$p_t$ distribution at the LHC.  In all cases, the SM is represented by 
the solid curve and the upper(lower) pairs of dotted (dashed, 
dash-dotted) curves correspond to $\kappa=0.5$ (-0.5), 0.25 (-0.25), 
0.125 (-0.125) respectively. From ref. \cite{rizzo-atqc1}}
\end{figure}

Rizzo examined the use of final state gluons as a probe for studying 
anomalous top-quark couplings at the NLC \cite{rizzo-atqc2}.  
He found that the rate and corresponding 
gluon jet energy distribution for the $e^+e^- \to t\bar{t}g$ are 
sensitive to the presence of anomalous couplings of the top to the 
photon and $Z$ at the production vertex as well as to the gluon 
itself. The sensitivity to anomalous gluon couplings 
is illustrated in Fig. 23 for a 1~TeV NLC. 
The resulting constraints are quite complementary to those 
obtained using other techniques.

\begin{figure}[h]
\leavevmode
\centerline{
\begin{turn}{-90}
\epsfig{file=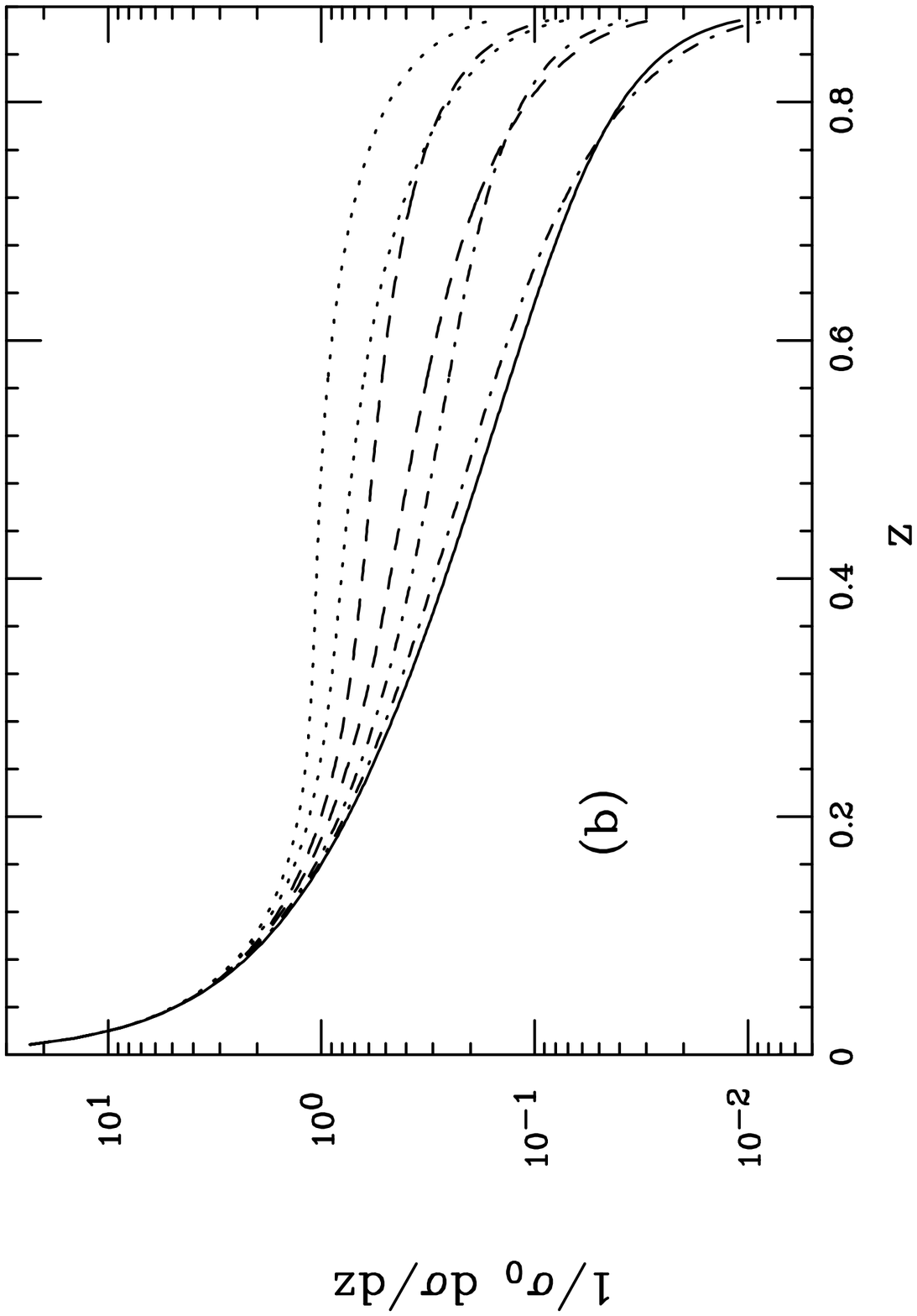,width=6.0cm,clip=}
\end{turn}
} 
\caption{Gluon jet energy spectrum assuming $\alpha_s=0.10$ for 
$m_t=175$~GeV at $\sqrt{s}=1$~TeV.  The upper (lower) dotted, dashed, 
and dot-dashed curves correspond to $\kappa$ values of 3(-3), 2(-2), 
and 1(-1) respectively while the solid curve is conventional QCD with 
$\kappa=0$. From ref. \cite{rizzo-atqc2}}
\end{figure}

\subsubsection{Excited Fermions}

Although it is expected that the first evidence for quark and/or 
lepton structure would arise from the effects of contact interactions 
direct evidence would be given by the observation of excited 
fermions.  Harris investigated the prospects for discovering an 
excited quark $u^*$ or $d^*$ decaying to dijets 
at hadron colliders \cite{harris}. 
The $qgq^*$ interaction is described by 
\begin{equation}
{\cal L}_{eff} =\frac{g}{2\Lambda} \bar{Q} \sigma^{\mu\nu} G_{\mu\nu} q
\end{equation}
where $Q$ represents the excited quark.
Harris considered the process $qg \to q^* \to qg$ via
a dijet resonance signal and included 
QCD background assuming an experimental dijet mass resolution of 10\%. 
The estimated $5\sigma$ discovery mass reach is 1.1~TeV at TeV33,
6.3~TeV at the LHC, 25~TeV at  a 50~TeV $pp$ collider, and
78~TeV at the 200~TeV PIPETRON collider.

\begin{table*}
\caption{New Interactions search reaches in TeV.  }
\begin{center}
\begin{tabular}{lcccccccccc}
\hline
\hline
Interaction & TeV & TeV33 & LHC & $pp$ & NLC & NLC & NLC & NLC & 
$\mu^+\mu^-$ & $\mu^+\mu^-$ \\
$\sqrt{s}$ (TeV) & 2  & 2 & 14 & 200 & 0.5 & 1.0 & 1.5 & 5.0 & 0.5 & 4.0 \\
$L$ fb$^{-1}$ & 0.11 & 30 & 100 &    & 50  & 200 & 200 & 1000 & 50 & 1000 \\
\hline
\multicolumn{11}{c}{Dimension 5 Anomalous Couplings} \\
\hline
$\Lambda$ & 2.8 & 3.5 & 17 & & $\sim 5$ & $\sim$ 10  & & & & \\
\hline
\multicolumn{11}{c}{Dimension 6 4-fermion Contact Interactions} \\
\hline
$\Lambda_{LL}(ee \to \mu \mu)$ & --- & --- & --- & --- 
	& 32 & 63 & 77 & 210 & 28 & 170 \\ 
$\Lambda_{LR}(ee \to \mu \mu)$ & --- & --- & --- & --- 
	& 29 & 57 & 70 & 190 & 25 & 150 \\ 
$\Lambda_{RL}(ee \to \mu \mu)$ & --- & --- & --- & --- 
	& 29 & 58 & 70 & 190 & 25 & 150 \\ 
$\Lambda_{RR}(ee \to \mu \mu)$ & --- & --- & --- & --- 
	& 31 & 62 & 76 & 210 & 27 & 160 \\ 
$\Lambda_{LL}(q\bar{q} \to \ell \ell)$ & 2.9 & 14 & --- & --- 
	&  & & & & & \\
$\Lambda_{LL}(q\bar{q} \to q \bar{q})$ & 1.6 & --- & 15 & --- 
	&  & & & & & \\
$\Lambda_{LL}(ee \to c \bar{c})$ & --- & --- & --- & --- 
	& 33 & 65 & 80 & 210 & 19 & 110 \\ 
$\Lambda_{LR}(ee \to c \bar{c})$ & --- & --- & --- & --- 
	& 28 & 57 & 69 & 190 & 16 & 92 \\ 
$\Lambda_{RL}(ee \to c \bar{c})$ & --- & --- & --- & --- 
	& 27 & 49 & 60 & 160 & 5.7 & 42 \\ 
$\Lambda_{RR}(ee \to c \bar{c})$ & --- & --- & --- & --- 
	& 34 & 62 & 76 & 210 & 15 & 90 \\ 
$\Lambda_{LL}(ee \to b \bar{b})$ & --- & --- & --- & --- 
	& 38 & 75 & 92 & 250 & 29 & 180 \\ 
$\Lambda_{LR}(ee \to b \bar{b})$ & --- & --- & --- & --- 
	& 30 & 61 & 77 & 210 & 22 & 140 \\ 
$\Lambda_{RL}(ee \to b \bar{b})$ & --- & --- & --- & --- 
	& 33 & 65 & 78 & 230 & 21 & 120 \\ 
$\Lambda_{RR}(ee \to b \bar{b})$ & --- & --- & --- & --- 
	& 33 & 67 & 82 & 250 & 20 & 120 \\ 
\hline
\multicolumn{11}{c}{Dimension 8 Contact Interactions} \\
\hline
$\Lambda^+(q\bar{q} \to \gamma\gamma)$ & 0.75 & --- & 2.8 & 23 &  & 
& & & & \\
$\Lambda^-(q\bar{q} \to \gamma\gamma)$ & 0.71 & --- & 2.9 & 16 &  & 
& & & & \\
\hline
\multicolumn{11}{c}{Discovery Reach for Excited Quarks} \\
	& 0.75 & 1.1 & 6.3 & 78 & --- & --- & --- & --- & --- & --- \\
\hline
\hline
\end{tabular}
\end{center}
\end{table*}

\section{Final Thoughts}

In this report we examined the potential of future collider facilities to 
study a broad range of new phenomena.  The range of physics topics that we
examined covered specific cases of new phenomena such as new gauge bosons, new 
fermions, and di-fermions, to more subtle hints of physics beyond the standard 
model via low energy effective operators which subsume the effects of new 
physics at a much higher energy scale. To deal with this abundance of 
possibilities, it seems to us that the most prudent approach in deciding upon
future facilities is to ensure that we are prepared for all of these
possibilities.  We should have the capability to
explore as many examples of new physics as our 
imagination can conceive and in as many processes as we can.  The ideal 
situation for this is to have hadron and lepton collider facilities with
comparable constituent center of mass energies.  As 
likely as not, when nature finally reveals her mysteries it will be 
totally unexpected.

At the time of writing, the two HERA experiments have indicated an
unexpected excess of events in Deep Inelastic Scattering at high $Q^2$.  
To interpret these observations (assuming they are not due to a statistical
fluctuation), results from a broad range of experiments have been found to be 
important: from the low energy 
precision atomic parity violation measurements to high energy results 
at the  1.8~TeV Tevatron $p\bar{p}$ collider and the 200 GeV LEP II
$e^+e^-$ collider.  This is a critical lesson; that although new physics 
may make its appearance at one facility, eg, the LHC, it is highly 
likely that measurements at a complementary facility such as the NLC may 
be crucial to understanding its significance, or visa versa.  To have
the best chance of discovering and understanding physics beyond the standard 
model requires  a diversified program of new facilities.

\section*{Acknowledgements}

S.G., J.H., and L.P. would like to thank the new phenomena working group
members and subgroup leaders for their enthusiasm and hard work.  
The authors also thank Robert Harris, Mathew Jones, and Tom Rizzo for 
their special assistance in the completion of this manuscript.
This
work is supported in part by the U.S. Department of Energy under contracts
DE-AC03-76SF00515 and W-31-109-ENG-38, and the Natural Sciences and 
Engineering Research Council of Canada.

%
\def\MPL #1 #2 #3 {Mod. Phys. Lett. {\bf#1},\ #2 (#3)}
\def\NPB #1 #2 #3 {Nucl. Phys. {\bf#1},\ #2 (#3)}
\def\PLB #1 #2 #3 {Phys. Lett. {\bf#1},\ #2 (#3)}
\def\PR #1 #2 #3 {Phys. Rep. {\bf#1},\ #2 (#3)}
\def\PRD #1 #2 #3 {Phys. Rev. {\bf#1},\ #2 (#3)}
\def\PRL #1 #2 #3 {Phys. Rev. Lett. {\bf#1},\ #2 (#3)}
\def\RMP #1 #2 #3 {Rev. Mod. Phys. {\bf#1},\ #2 (#3)}
\def\ZPC #1 #2 #3 {Z. Phys. {\bf#1},\ #2 (#3)}

\end{document}